\documentclass[prc,twocolumn,superscriptaddress]{revtex4-1}
\usepackage{epsfig}
\usepackage{graphicx}
\usepackage{palatino}
\usepackage[normalem]{ulem}
\usepackage[english]{babel}
\usepackage{hyphenat}
\usepackage{amsmath}
\usepackage{amssymb}
\usepackage{slashed}
\usepackage{epstopdf}
\usepackage{xcolor}
\usepackage{booktabs}
\definecolor{lcolor}{rgb}{0.,0.0,0.}
\definecolor{citcolor}{rgb}{0,0.,0.5}
\usepackage[breaklinks,colorlinks,urlcolor=blue,citecolor=blue,linkcolor=blue]{hyperref}
\usepackage{multirow}
\usepackage{ltablex}
\newcommand{\beq}{\begin{equation}}
\newcommand{\eeq}{\end{equation}}
\newcommand{\bea}{\begin{eqnarray}}
\newcommand{\eea}{\end{eqnarray}}

\def\dd{{\rm d}}

\newcommand{\bem}{\begin{multline}}
\newcommand{\eem}{\end{multline}}
\newcommand{\beg}{\begin{gather}}
\newcommand{\eeg}{\end{gather}}

\newcommand{\ben}{\begin{eqnarray*}}
\newcommand{\een}{\end{eqnarray*}}

\newcommand{\eq}[1]{\begin{align}#1\end{align}}
\newcommand{\bal}{\begin{align}}
\newcommand{\eal}{\begin{align}}

\newcommand{\eqn}[1]{Eq.~\eqref{#1}}

\newcommand{\secn}[1]{Section~1}
\newcommand{\appn}[1]{Appendix~1}

\long\def\comment#1{ }

\def\and{\quad\text{and}\quad}

\def\0{{\boldsymbol 0}}

\def\max{{\rm max}}

\begin{document}

\title{Tagging boosted hadronic objects with dynamical grooming}
\author{Yacine Mehtar-Tani}
\email[]{mehtartani@bnl.gov}
\affiliation{Physics Department, Brookhaven National Laboratory, Upton, NY 11973, USA}
\affiliation{RIKEN BNL Research Center, Brookhaven National Laboratory, Upton, NY 11973, USA}
\author{Alba Soto-Ontoso}
\email[]{ontoso@bnl.gov}
\affiliation{Physics Department, Brookhaven National Laboratory, Upton, NY 11973, USA}
\author{Konrad Tywoniuk}
\email[]{konrad.tywoniuk@uib.no}
\affiliation{Department of Physics and Technology, University of Bergen, 5007 Bergen, Norway}

\begin{abstract}
We evaluate the phenomenological applicability of the dynamical grooming technique, introduced in \cite{Mehtar-Tani:2019rrk}, to boosted W and top tagging at LHC conditions. An extension of our method intended for multi-prong decays with an internal mass scale, such as the top quark decay, is presented. First, we tackle the reconstruction of the mass distribution of W and top jets quantifying the smearing due to pileup. When compared to state-of-the-art grooming algorithms like SoftDrop and its recursive version, dynamical grooming shows an enhanced resilience to background fluctuations. In addition, we asses the discriminating power of dynamical grooming to distinguish W (top) jets from QCD ones by performing a two-step analysis: introduce a cut on the groomed mass around the W (top) mass peak followed by a restriction on the N-subjettinnes ratio $\tau_{21}$ ($\tau_{32}$). For W jets, the out-of-the-box version of dynamical grooming, free of ad-hoc parameters, results into a comparable performance to SoftDrop. Regarding the top tagger efficiency, $3$-prong dynamical grooming, in spite of its simplicity, presents better performance than SoftDrop and similar results to Recursive SoftDrop.
\end{abstract}

\maketitle

\section{Introduction}
\label{sec:intro}
The forthcoming high-luminosity phase of the Large Hadron Collider (HL-LHC) will pursue the discovery of new resonances beyond the Standard Model (BSM) of particle physics with masses around the TeV scale. Experimentally, the observation of heavy particles using their hadronic decay channels involves events with multi-jet signatures and cumbersome topologies in extreme regions of phase space. On many occasions, these particles are produced in a highly boosted regime and, consequently, the opening angle between their decay products is small. As a result, they end up reconstructed as a single large radius jet that: i) hides the rich multi-prong structure of these decay channels and ii) can be easily misidentified as a \textit{conventional} jet arising from a quark or gluon splitting due to the QCD collinear singularity. The relevant question on how to experimentally distinguish a \textit{background} QCD jet from a new physics signal is addressed by so-called ``jet taggers'' that aim at exploiting the fundamental differences in the radiation patterns.

During the last decade, the landscape of taggers has significantly expanded, as summarized in~\cite{Larkoski:2017jix, Marzani:2019hun, Bendavid:2018nar}. These developments have occurred hand-in-hand with the maturing of theoretically motivated jet substructure techniques~\cite{Asquith:2018igt, Kaplan:2008ie, Thaler:2011gf,Dasgupta:2013ihk,Dasgupta:2013via,Napoletano:2018ohv} and machine learning methods~\cite{Kasieczka:2019dbj,Collins:2019jip,Andreassen:2020nkr,Nachman:2020lpy}. In what follows, we will focus on experimental searches where the properties of the final state, such as the number of jets and their prongs, are known and, therefore, semi-analytical approaches to jet tagging are applicable and competitive with machine learning. 

Generically, a tagging algorithm starts by minimizing the impact of non-perturbative contributions, like underlying event, hadronization and pileup, to the reconstructed jet through a grooming algorithm~\cite{Chien:2019osu,Carrazza:2019efs,Dreyer:2018tjj,Larkoski:2014wba,Mehtar-Tani:2019rrk,Butterworth:2008iy}. This step allows to sharpen the mass spectrum of the jet coming from the boosted hadronic object, that forms a narrow distribution in the absence of soft contamination, and, therefore, optimize the signal-to-background ratio in the mass window around the resonance peak~\cite{Kang:2018jwa,Marzani:2017kqd}. Grooming also helps to identify the number of hard prongs in the jet, e.g. three for a top decay, although other techniques such as pruning~\cite{PhysRevD.81.094023} or trimming~\cite{Krohn:2009th} might be used for this purpose. Once the spurious radiation has been groomed away, additional cuts on so-called jet shape variables, such as the N-subjettinnes ratios $\tau_{ij}$~\cite{Thaler:2010tr,Thaler:2011gf}or energy correlation functions~\cite{Larkoski:2013eya,Komiske:2017aww}, are performed in order to isolate the relevant corners of the radiation phase space and enhance the QCD background rejection power. Making use of these techniques, ATLAS and CMS have reported a plethora of experimental results that include analyses of Standard Model particles~\cite{Aad:2015rpa, Sirunyan:2019qia} and new physic searches~\cite{Schramm:2018uyb}. 

As in many other particle physics areas, the debate on whether theoretically sound and easy to implement techniques should be preferred over high-level algorithms that involve machine-learning methods is pertinent and timely in the context of jet tagging. In this paper we opt for the former option and propose an economic, free of ad-hoc parameters, yet well performing two and three prong tagger where the underlying dynamics can be analytically pinned down. The method consists of two steps: 
\begin{enumerate}
\item{}We apply dynamical grooming (DyG) either in its original guise~\cite{Mehtar-Tani:2019rrk} or with a multi-prong extension ($3$-Prong DyG), that will be introduced below. This method selects the hardest splitting in an angular ordered shower, given the definition of \textit{hardness}, defined through the variable $\kappa^{(a)}$ (cf. \eqn{eq:hadrness}). We explore three cases to characterize the hardness of a splitting: the momentum sharing fraction ($z$Drop), the relative transverse momentum ($k_T$Drop) and the virtuality or its inverse, formation time (TimeDrop). Once the hardest splitting has been identified, one discards all emissions taking place at larger angles in the reclustering sequence. Only those jets whose dynamically groomed mass is contained in the interval of a given width $\delta M$ around the resonance peak $M_X$, i.e. [$M_X-\delta M, M_X+\delta M$], are accepted. 

\item{}Next, we follow previous works in the literature, e.g.~\cite{Dreyer:2018tjj}, and, using the dynamically groomed jet, perform a cut on the relevant $N$-subjettiness ratio, i.e. the one that is clearly different in signal and QCD jets. A jet is tagged as signal if, besides satisfying the mass constraint, its jet shape satisfies $\tau_{ij}\!<\!\tau_{\rm cut}$.
\end{enumerate}

Notice that, as mentioned, the second step in the tagging process is rather standard and thus, the main novelty of this work is to use dynamical grooming to simultaneously identify the number of hard prongs and groom away non-perturbative radiation. This idea is strongly motivated by a previous publication~\cite{Mehtar-Tani:2019rrk}, where we explored the properties of the tagged splitting in QCD jets finding: i) the proposed analytic framework, based on vetoed showers, qualitatively agrees with Monte-Carlo simulations, ii) indications of a remarkable resilience of dynamically groomed observables to non-perturbative effects, including both hadronization and underlying event. 

This manuscript is organized as follows. In Sec.~\ref{sec:grooming}, we concisely summarize state-of-the art grooming techniques utilized along this manuscript such as SoftDrop~\cite{Larkoski:2014wba}, its recursive version~\cite{Dreyer:2018tjj}, and dynamical grooming~\cite{Mehtar-Tani:2019rrk}. Then, we study the groomed mass spectrum for W jets including the effect of pileup in current and future high-luminosity conditions at the LHC, in Sec.~\ref{sec:w}. In order to asses the tagging performance of the proposed method we present the tagging efficiency of W jets against the QCD background rate in Sec.~\ref{sec:wtag}. Next, in Sec.~\ref{sec:top} we analyze jets arising from top decays. To accommodate three-prong topologies in our framework, we introduce $3$-Prong DyG in Sec.~\ref{sec:3prongDyG}. The top mass distribution with and without pileup together with a study on boosted top tagging are presented in Sec~\ref{sec:topmass}, Sec.~\ref{sec:toptag}, respectively. We end with a discussion on our findings in Sec.~\ref{sec:conclusions}. 

\section{Grooming techniques}
\label{sec:grooming}
For the sake of completeness, we briefly revisit the grooming techniques used in this paper in an algorithmic fashion. We refer the reader to the original publications for a more detailed explanation of each of the methods.

\subsection{SoftDrop}
Over the last five years, SoftDrop (SD)~\cite{Larkoski:2014wba}, an extension of the modified Mass-Drop Tagger (mMDT) \cite{Dasgupta:2013ihk}, has become a preferred choice for theoretical and experimental analyses of jet substructure, see~\cite{Aad:2019vyi} and~\cite{Adam:2020kug} for recent results at LHC and RHIC, respectively. This success is rooted in the possibility to systematically compute SoftDrop observables within perturbative QCD~\cite{Kang:2019prh,Marzani:2017kqd,Kang:2018jwa} together with its relatively simple algorithmic implementation that proceeds as follows:
\begin{itemize}
\item Re-cluster the jet candidate, $j$, with the Cambridge/Aachen (C/A) algorithm~\cite{Dokshitzer:1997in}.
\item Undo the last clustering step and compute the momentum sharing fraction, $z$ among the subjets ($j_1,j_2$)
\eq{z=\displaystyle\frac{{\rm{min}}(p_{T,1},p_{T,2})}{p_{T,1}+p_{T,2}}.
}
and their opening angle $\theta$.
\item If the SoftDrop condition is satisfied i.e.
\eq{z>z_{{\rm{cut}}} \theta^\beta
\label{eq:sdcondition}
}
the algorithm ends and $j$ is returned as the SoftDrop jet. 
\item If not, remove the softest subjet and repeat the de-clustering process on the hardest branch. 
\end{itemize}

In Eq.~\eqref{eq:sdcondition}, ($z_{{\rm{cut}}}, \beta$) are free parameters that, in principle, entail a certain degree of flexibility to select splittings from different kinematic regions (where $\beta\!=\!0$ is equivalent to mMDT). However, this is a manifest disadvantage when applying the method to experimental data as there is no optimal pair of values a priori. In reality, they have to be determined  on an observable-by-observable basis and might depend on the properties of the jet such as its $p_T$, radius or mass, as well as the background. This fact naturally introduces a Monte-Carlo dependence on experimental analyses that contributes to the systematic error of the measurement. 

\subsection{Recursive SoftDrop}
Recently, an extension of the SoftDrop algorithm has been proposed~\cite{Dreyer:2018tjj}, called recursive SoftDrop (RSD). The idea is to search along the full clustering tree for the $N$ splittings that satisfy the SoftDrop condition, see Eq.~\eqref{eq:sdcondition}, such that the groomed jet is a collection of $N\!+\!1$ prongs (RSD$_{N}$). A more aggressive grooming can be achieved by recursing along the complete clustering sequence (RSD$_{\infty}$). The main performance improvement with respect to traditional SoftDrop occur in scenarios with large pileup conditions (although it still needs to be complemented with a background subtraction method) or multi-prong boosted objects, or both. This method generalizes iterative SoftDrop~\cite{Frye:2017yrw}, where only the nodes that satisfy the SoftDrop condition along the primary Lund plane are taken into account. Although iterative SoftDrop could be used as a grooming method, the associated class of counting observables related to the selected nodes, such as their multiplicity $n_{{\rm{SD}}}$, have attracted more attention, specially in the heavy-ion context, both experimentally~\cite{Acharya:2019djg} and theoretically~\cite{Casalderrey-Solana:2019ubu, Caucal:2019uvr}.

Rather than performing an optimization procedure, we stick to the recommendations by the authors in~\cite{Dreyer:2018tjj}  for both SD and RSD and set ($z_{{\rm{cut}}}\!=\!0.05, \beta\!=\!1$) in Eq.~\eqref{eq:sdcondition}. Further, as we only explore either two (W) or three (top) prong decays we restrict ourselves to SD, RSD$_{2}$ and RSD$_{\infty}$. 

\subsection{Dynamical grooming}
\label{sec:dyg}
Aiming at reducing the number of ad-hoc parameters in current grooming algorithms and taking into account the jet-by-jet fluctuating nature of jet substructure, an algorithm with dynamically generated grooming conditions has been recently proposed~\cite{Mehtar-Tani:2019rrk}.

In order to dynamically groom (DyG) a jet, $j$, the next steps have to be followed:
\begin{itemize}
\item Re-cluster $j$ with C/A.
\item Find the hardest branch in the primary Lund plane~\cite{Andersson:1988gp,Dreyer:2018nbf}, i.e. the splitting that satisfies
\begin{align}\label{eq:hadrness}
\kappa^{(a)} =\frac{1}{p_T}\,\,\underset{i\in\, \text{LP}}{\max}\left[z_i(1-z_i) \, p_{T,i}\,\left(\frac{\theta_i}{R}\right)^a  \right]\,,
\end{align}
where $a$ is a continuous free parameter that satisfies $a\!>\!0$ to guarantee IRC safety~\cite{Mehtar-Tani:2019rrk}. 
\item Drop all branches at larger angles, that is, prior in the C/A sequence. 
\end{itemize}

As already introduced, the values of $a$ explored in this paper are: $a\!=\!0.1$ (zDrop), $a\!=\!1$ ($k_T$Drop) and $a\!=\!2$ (TimeDrop). The procedure just outlined can be thought of as removing soft radiation sensitive to the total color charge of the jet. A connection between the DyG parameter $a$ and ($z_{{\rm{cut}}}, \beta$) in SoftDrop can be established. Notice that the lower the value of $a$, the more aggressive the grooming becomes, just like reducing $\beta$ in the SD condition. Further, DyG dynamically generates a value of $z_{{\rm{cut}}}$ that scales as $z_{{\rm{cut}}}\sim e^{-\sqrt{a/\alpha_s}}$ at leading-log accuracy~\cite{Mehtar-Tani:2019rrk}.

\section{W jets}
\label{sec:wjets}
Having exposed the basic methodology used in this work, let us now turn to the first application, namely to reconstruct the $W$ boson mass and width parameters in the boosted regime, i.e. from two-pronged jets.

To numerically generate our data samples we choose {\tt{Pythia 8.235 tune 14}}~\cite{Sjostrand:2007gs} to simulate 25k p+p collisions at $13$~TeV. All particles in each event are clustered into anti-$k_T$ jets~\cite{Cacciari:2008gp} with $R\!=\!0.8$ and re-clustered with Cambridge/Aachen using {\tt{FastJet 3.1}}~\cite{Cacciari:2011ma}. The analysis is performed on jets with $p_{T}\! >\! 450$~GeV and $|\eta|\!<\!3$. To simulate pileup, we embed the hard process into $n_{{\rm{PU}}}$ minimum bias events. Then, the analysis is performed on reconstructed jets of the full event geometrically matched to jets with signal particles only in order to avoid the well known influence of pileup on jet finding algorithms. Throughout this work, detector effects are not included.

\subsection{W mass distribution}
\label{sec:w}
\begin{figure}
\includegraphics[width=\columnwidth]{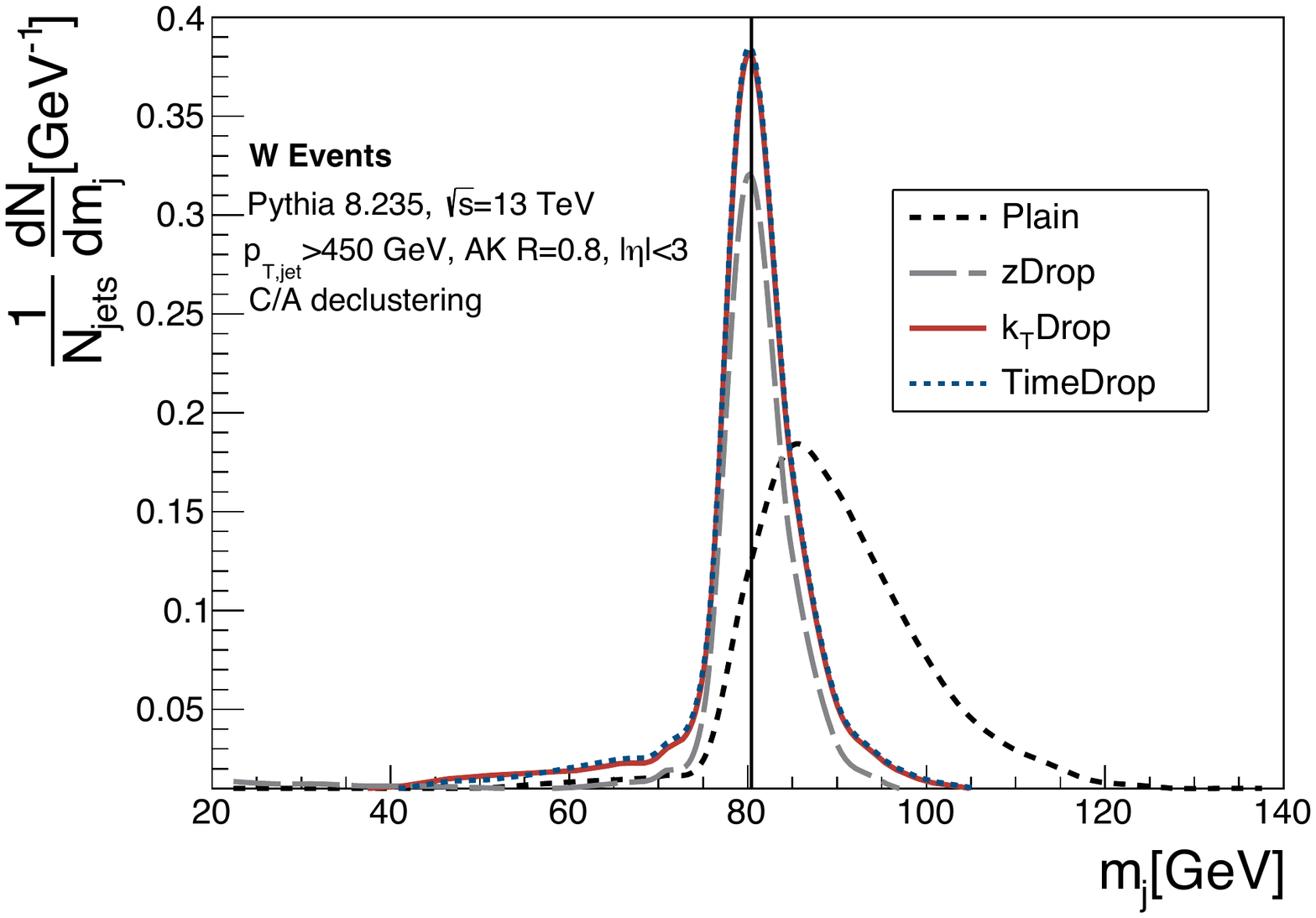}
\caption{The dynamically groomed mass distribution of W jets for $z$Drop (dashed, gray), $k_T$Drop (solid, red) and TimeDrop (dotted, blue). For completeness, the vertical black line indicates the W boson mass and the un-groomed (plain) distribution is given by the dashed, black line.}.
\label{fig:mass-wdyg}
\end{figure}

\begin{figure*}
\includegraphics[width=\textwidth]{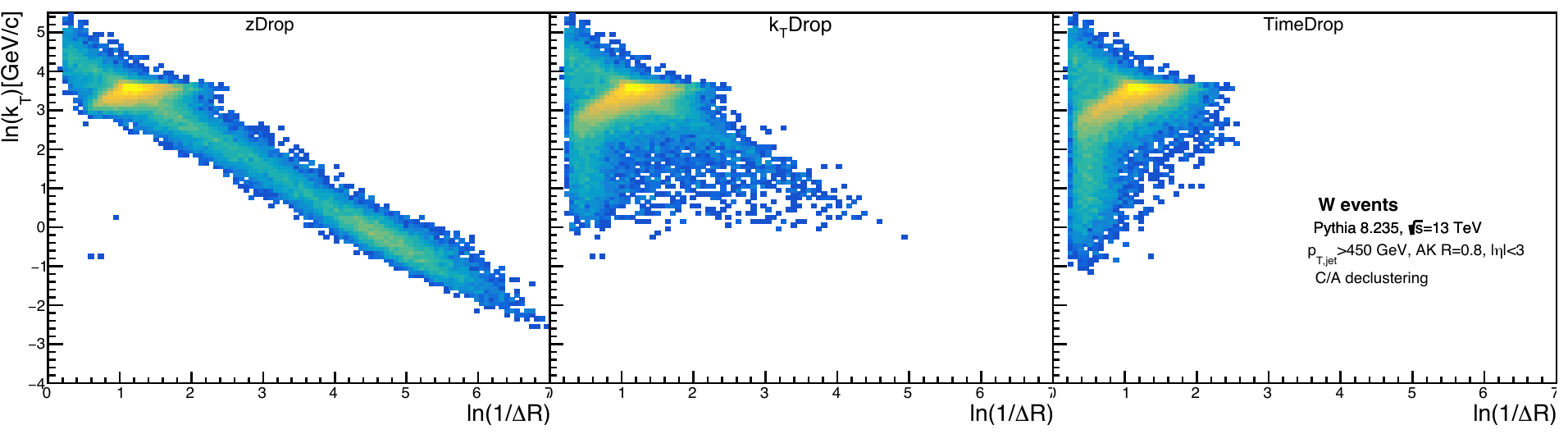}
\caption{Primary Lund planes for the tagged emissions of the dynamical grooming family in W events.}
\label{fig:lp-wdyg}
\end{figure*}

In this case, we generate $p\!+\!p\!\rightarrow\!WW$ events with the electroweak boson decaying hadronically $W\!\rightarrow\!q\bar{q}$. The jet mass distribution of these events is shown in Fig.~\ref{fig:mass-wdyg}. We notice that TimeDrop and $k_T$Drop result into rather similar distributions, meaning that these values of $a$ in Eq.~\eqref{eq:hadrness} often result into tagging the same splitting. This idea is confirmed by the primary Lund planes displayed in Fig.~\ref{fig:lp-wdyg} where the characteristic shape of the most densely populated area is a result of the intrinsic scale in this scenario, i.e. the W mass. We observe that most of the splittings selected by TimeDrop coincide with those of $k_T$Drop. Although not shown in Fig.~\ref{fig:lp-wdyg}, these splittings are also captured by SoftDrop. This is expected as long as the W decay into the $q\bar{q}$ is sufficiently hard and not too collinear. In contrast, the resulting Lund plane from $z$Drop reveals a significant sensitivity to small angle radiation causing the mass peak shift towards larger values observed in Fig.~\ref{fig:mass-wdyg}. 

Besides underlying event and hadronization, responsible for the mass distribution smearing on Fig.~\ref{fig:mass-wdyg},  another contribution to the distortion of the W mass spectrum at the LHC is pileup. To mimic this contamination, we embed W events into a number $n_{{\rm{PU}}}$ of minimum bias events, dominated by soft QCD processes. For current LHC conditions $n_{{\rm{PU}}}$ is set to 60 while the high-luminosity phase is modeled with $n_{{\rm{PU}}}\!=\!200$, as estimated in~\cite{Atlas:2019qfx}. In the presence of pileup, we have checked that grooming is not enough to remove the spurious radiation and a dedicated background subtraction technique has to be included in the analysis. We use Constituent Subtraction (CS)~\cite{Berta:2014eza} with $\alpha\!=\!1$. For simplicity, the area-median~\cite{Cacciari:2007fd} is used as the input $p_T$-background estimator although the overall performance could be improved by using more precise background estimators such as~\cite{Haake:2018hqn,Yacine:2019ycj}. Once CS has been applied on the clustered jets, they are groomed. 

To quantify the performance of different grooming methods in describing the mass spectrum, we follow the strategy proposed in~\cite{Dreyer:2018tjj}. That is, we find the smallest interval $[m_{j,\rm min},m_{j,\rm max}]$ that contains $40\%$ of the total number of events. Then, the position of the mass peak $M_X$ is defined as the median of this interval while the mass resolution is given by $\Gamma_X\!=\!m_{j,\rm max}\!-\!m_{j,\rm min}$. 

\begin{figure*}
\includegraphics[scale=0.45]{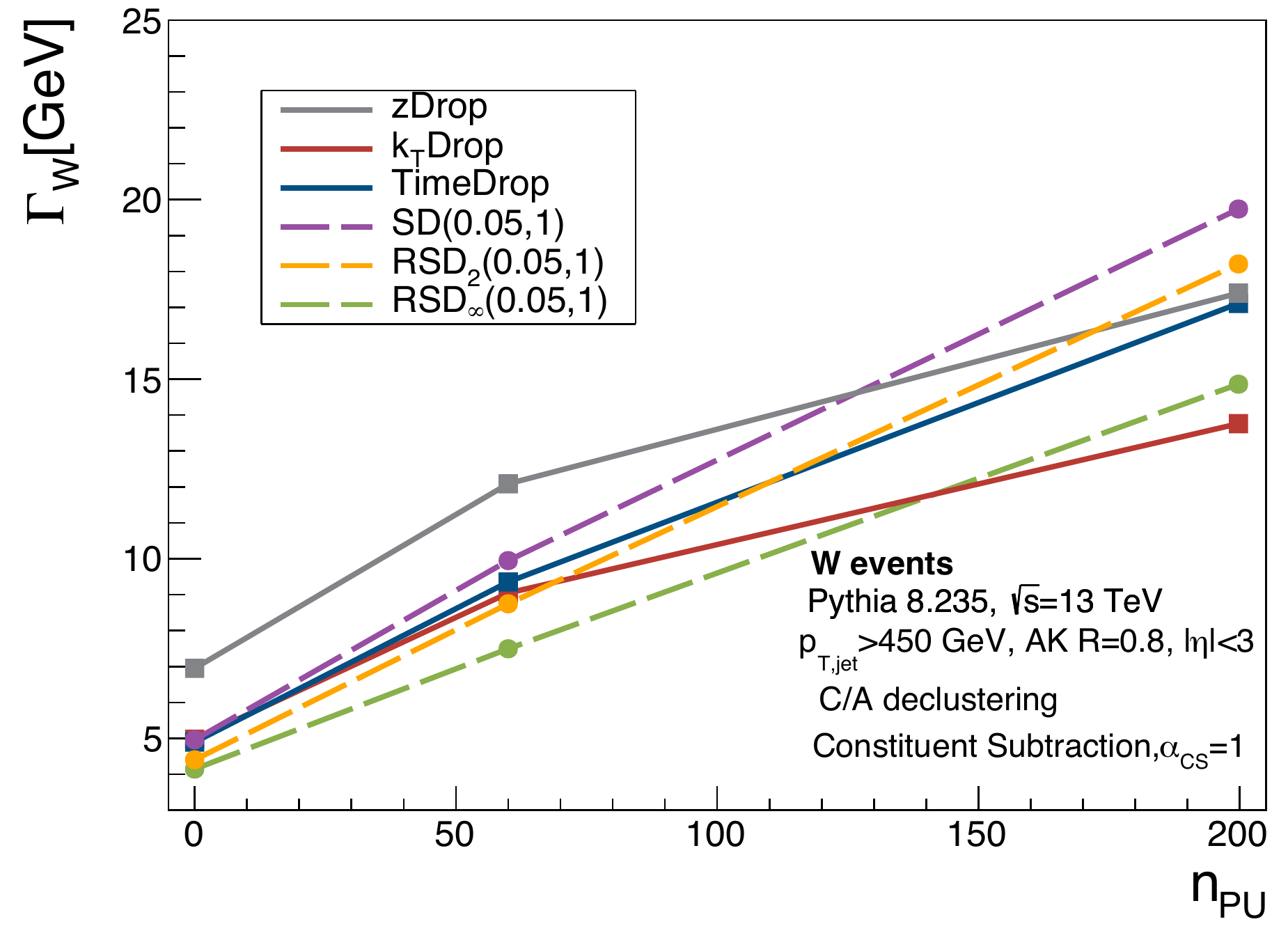}\quad\includegraphics[scale=0.45]{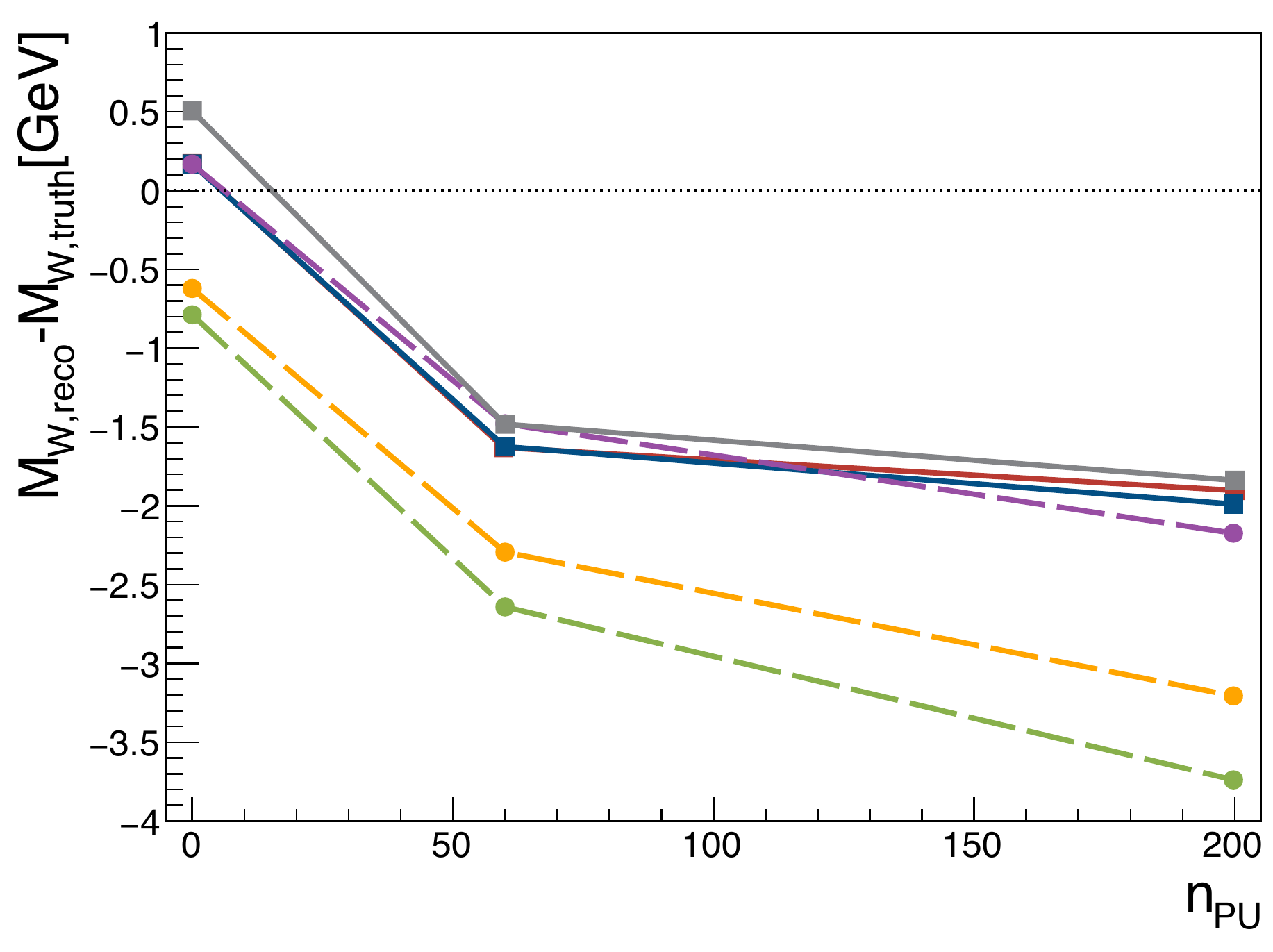}
\caption{Impact of pileup on the values of the W mass width (left) and its peak position with respect to the W boson mass $M_{W,{\rm{truth}}}$ (right) for different grooming methods at current LHC conditions ($n_{\rm_{PU}}\!=\!60$) and future HL-LHC ($n_{{\rm{PU}}}\!=\!200$).}
\label{fig:mass-w-comparison-wpu}
\end{figure*}

The values of the mass peak position $M_W$ and the width of the distribution $\Gamma_W$ are shown in Fig.~\ref{fig:mass-w-comparison-wpu} for no pileup, current LHC and HL-LHC. In the absence of pileup, we observe that TimeDrop and $k_T$Drop deliver similar results to SD without the need to tune any parameter, a crucial aspect regarding the applicability of the method to raw experimental data. In all three cases, the mass peak is accurately described with a resolution of less than $5$~GeV, that is, $\Gamma_W\sim 6\% \,M_W$. While recursively applying SoftDrop through the C/A tree further improves the mass resolution, the peak shifts towards smaller values indicating that too much radiation was groomed away for this choice of ($z_{{\rm{cut}}}, \beta$). 

Turning to the pileup contaminated scenarios, we have first checked that grooming is necessary in this environment as the plain $M_W$, even after applying the background subtraction technique, are $10\!-\!20$~GeV off the resonance value. As expected, we observe that all methods result in a broader mass distribution with respect to the no pileup case, but also in an underestimation of the W mass. For $n_{\rm_{PU}}\!=\!60$, $k_T$Drop and TimeDrop performance is in the ballpark of the SoftDrop family indicating their robustness against this type of contamination. In the extreme regime of HL-LHC, there is a larger spread among the results of different grooming strategies. In particular, the different performance of RSD$_{\infty}$ with respect to its non-recursive counterpart results into a $5$~GeV smaller $\Gamma_W$, but at the cost of having a $\!\sim\!4$~GeV shift on $M_W$. Nevertheless, the smallest mass resolution is achieved by $k_T$Drop together with the closest $M_W$ to the truth value, although still $1.5$~GeV off.  It is worth noting that while DyG and SD present similar performance for $n_{\rm_{PU}}\!=\!60$, in the larger pileup case $n_{\rm_{PU}}\!=\!200$, the mass width $\Gamma_W$ for $k_T$Drop (for instance) is 20 to 30 \% smaller than that resulting from RSD$_2$ and SD, respectively, as can be observed in Fig.~\ref{fig:mass-w-comparison-wpu}. At the same time, the mass peak remains within $2$~GeV of the resonance value, in sharp contrast with RSD$_2$ and RSD$_\infty$. These two facts suggest a better performance of jet-by-jet dynamical grooming compared to SD with increasing pileup.

\subsection{Boosted W tagging}
\label{sec:wtag}
\begin{figure}
\includegraphics[width=\columnwidth]{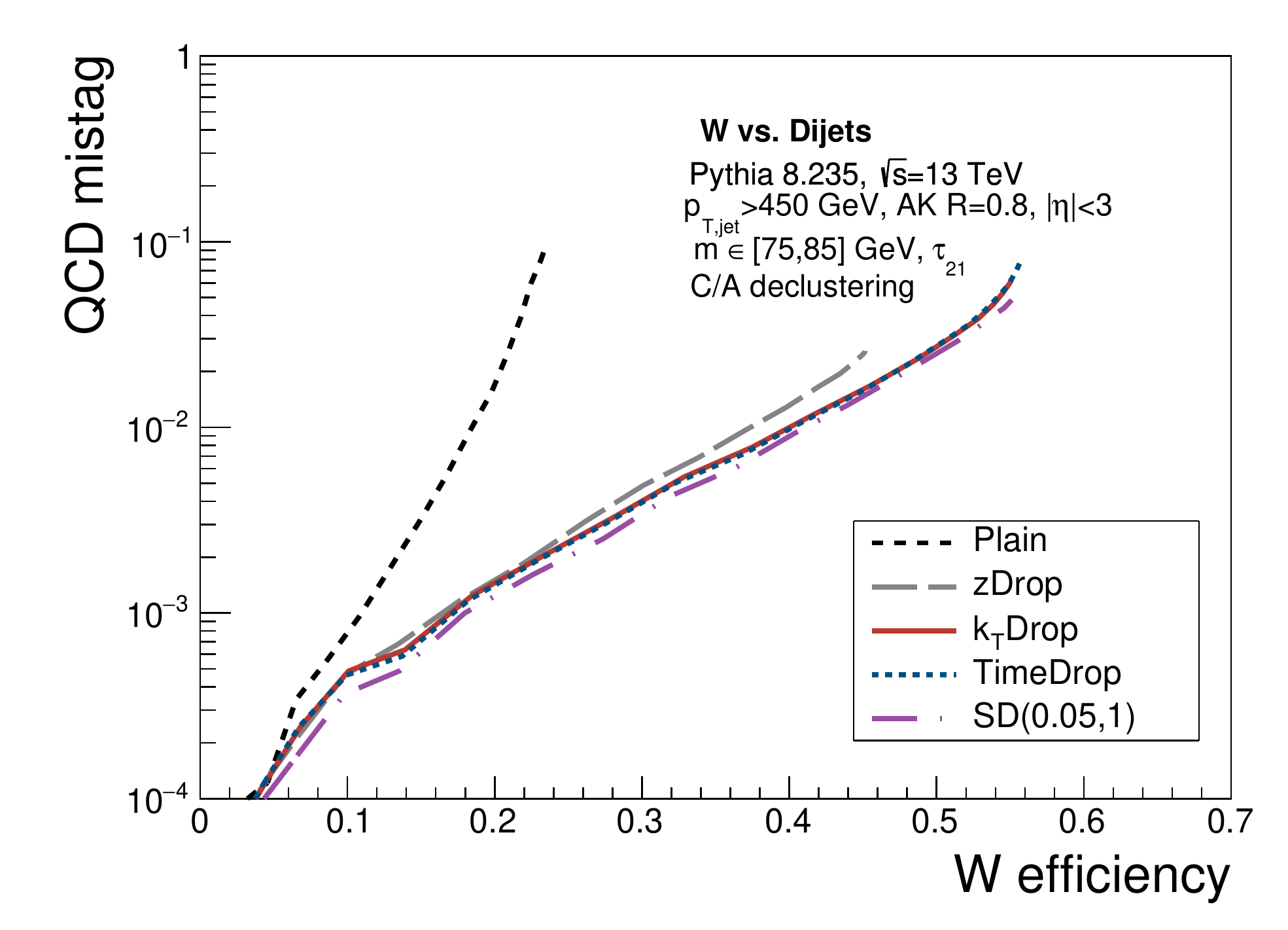}
\caption{QCD mistag rate as a function of the W tagging efficiency without grooming (solid, black) and with different grooming techniques: SoftDrop (dotted-dashed, purple), zDrop (dashed, gray), $k_T$Drop (solid, red) and TimeDrop (dotted, blue).}
\label{fig:w-vs-dijets}
\end{figure}

In the previous Section, we have studied how grooming helps to sharpen the W mass spectrum.  We now turn to addressing the problem of distinguishing a jet arising from the electroweak decay of a W boson from the fragmentation of a quark/gluon in the absence of pileup. 

A selection based on the (un)groomed mass is insufficient as QCD jets are copious at the LHC and their mass distribution substantially overlaps with the one of W jets around the resonance peak. To enhance the discriminating power, it has been proposed to add a second step on the tagging process that consists on restricting the radiation pattern by doing cuts on jet shapes. For simplicity, we are going to focus on the use of N-subjettiness. As a remainder, the definition is \cite{Thaler:2010tr,Thaler:2011gf}
\beq
\tau_{N}^{\beta}=\displaystyle\sum_{i\in{\rm jet}}z_{i}{\rm{min}}(\theta_{i,a_1}^\beta,\ldots,\theta_{i,a_N}^\beta)
\eeq
where we choose $a_i$ to be generalized $k_{T}$-axes and $\beta$ is set to $2$. The basic idea is that for a jet with $N$ prongs, one expects $\tau_1,\ldots,\tau_{N}$ to be large and $\tau_{>N}$ to be small. Therefore, in our context, where we want to distinguish W from QCD jets, the optimal choice is the ratio
\beq
\tau_{21}=\displaystyle\frac{\tau_2}{\tau_1}
\eeq
that would be larger for QCD jets than for W's. Then, as introduced in Sec.~\ref{sec:intro}, a jet would be tagged as sourced by a W decay if, after grooming, $\tau_{21}\!<\!\tau_{{\rm cut}}$ and $|m_{{\rm jet}}\!-\!M_{W}|\!<\!\delta M$. 

In order to evaluate the performance of a given strategy we compute $\varepsilon^{\rm{QCD}}_B$, as
\beq
\label{eq:efficiency}
\varepsilon^{\rm{QCD}}_B([m_{\rm{min}}, m_{\rm {max}}]; \tau_{\rm cut})
=\displaystyle\frac{\displaystyle\int_{m_{\rm{min}}}^{m_{\rm{max}}} \dd m\displaystyle\frac{1}{\sigma}\displaystyle\frac{\dd\sigma}{\dd m}\Big\vert^{\rm{QCD}}_{\tau_{21}<\tau_{{\rm cut}}}}{\displaystyle\int_{0}^{\infty} \dd m\displaystyle\frac{1}{\sigma}\displaystyle\frac{\dd\sigma}{\dd m}\Big\vert^{\rm{QCD}}}.
\eeq

That is, one counts the number of jets in the QCD sample whose (un-)groomed mass is in the range $[m_{\rm{min}}, m_{\rm {max}}]$ and satisfy $\tau_{21}\!<\!\tau_{{\rm cut}}$. Then, we normalize by the total number of jets. Repeating the same process for the W sample, we obtain the tagging efficiency for a given value of $\tau_{{\rm cut}}$. By varying the value of $\tau_{{\rm cut}}$ in Eq.~\eqref{eq:efficiency}, a receiver operator curve (ROC) might be generated. In this representation, the larger the W efficiency is while keeping the QCD mistag rate small, the better performance a tagger has. 

Such ROC curves are visualized in Fig.~\ref{fig:w-vs-dijets}, where the QCD mistag rate is plotted as a function of the W efficiency. Notice that we impose a rather tight interval on the (un)groomed mass. We confirm the results of the previous Section and observe that $k_T$Drop and TimeDrop result into a similar W tagging performance as SoftDrop for W tagging efficiencies above 40$\%$. Although not shown in Fig.~\ref{fig:w-vs-dijets}, we find similar tagging results for RSD$_{2}$ and RSD$_{\infty}$. Overall, both (R)SD and DyG result into a neat improvement with respect to the plain case. Given all the results displayed in this Section, we confirm both $k_T$Drop and TimeDrop as reliable techniques for two-pronged boosted object identification on experimental data recorded at the LHC.

\section{Top jets}
\label{sec:top}
Most new physics scenarios include heavy particles that couple and decay to top quarks. Thus, an efficient method to identify jets arising from top quarks is crucial for the physics prospects of the HL-LHC. The top quark predominantly decays into a $b$ quark and a W boson that subsequently decays into a pair of light quarks. Besides the problem of $b$-tagging, these three jets are almost indistinguishable in a highly boosted scenario and end up forming part of a single large-$R$ jet. This complicated topology poses a challenge for the dynamical grooming technique and calls for an extension of the method to handle 3-prong decays, as described below.

\subsection{3-prong DyG}
\label{sec:3prongDyG}
\begin{figure}
\includegraphics[width=0.75\columnwidth]{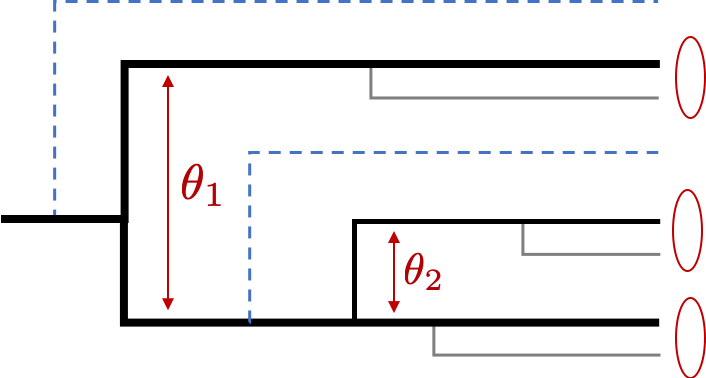}
\caption{3-prong dynamical grooming applied to an angular ordered tree. The blue-dashed lines represent the groomed branches and the angles $\theta_1$, $\theta_2$ are defined in the text.}
\label{fig:3pdyg}
\end{figure}
A crucial aspect in the line of reasoning of DyG is the assumption of angular ordering in the shower which to leading logarithmic accuracy insures that the {\it hardness} variable decreases along the shower. While this is certainly the case in QCD jets, there are other scenarios such as top quark decay, where angular ordering is not a particularly useful guiding principle for the hard event formation. To circumvent this problem and capture the three-prong topology of the top decay, we propose the following modification of the original procedure described in Section~\ref{sec:grooming}:   
\begin{itemize}
\item Re-cluster $j$ with C/A.
\item First, find the hardest $j_1$, with associated splitting angle $\theta_\text{leading}$. 
\item Next, identify the next-to-hardest $j_2$ splitting, with associated splitting angle $\theta_\text{sub-leading}$ located either
\begin{itemize}
\item on the primary Lund plane of  $j_1$ (in this case the next-to-hardest can also occur \textit{before} the hardest splitting, i.e. $\theta_\text{sub-leading} > \theta_\text{leading}$), or 
\item on the secondary Lund plane associated to the softest daughter particle.
\end{itemize}
\item Denoting $\theta_1\!\equiv\!\max (\theta_\text{leading}, \theta_\text{sub-leading})$ and $\theta_2\!\equiv\!\min(\theta_\text{leading}, \theta_\text{sub-leading})$, drop all $i$ branches with 
\begin{itemize}
\item $\theta_i\!>\!\theta_1$ along the primary branch.
\item $\theta_1\!>\!\theta_i\!>\!\theta_2$ along the branch where $j_2$ belongs.
\end{itemize}
\end{itemize}
A sketch of this algorithm, dubbed $3$-Prong DyG, is shown in Fig.~\ref{fig:3pdyg}. 
Determining the next-to-hardest splitting constitutes the minimal extension of the method to capture the third prong. Notice that the key input is the number of prongs and thus this idea could be generalized for an arbitrary number of hard prongs in the final state jet, $n$-Prong DyG. In particular, $4$-Prong DyG would be relevant, for example, in Higgs searches, another key sector for SM measurements and new physics \cite{Butterworth:2008iy}. This possibility will be considered in future works.

\subsection{Top mass distribution}
\label{sec:topmass}
\begin{figure}
\includegraphics[width=\columnwidth]{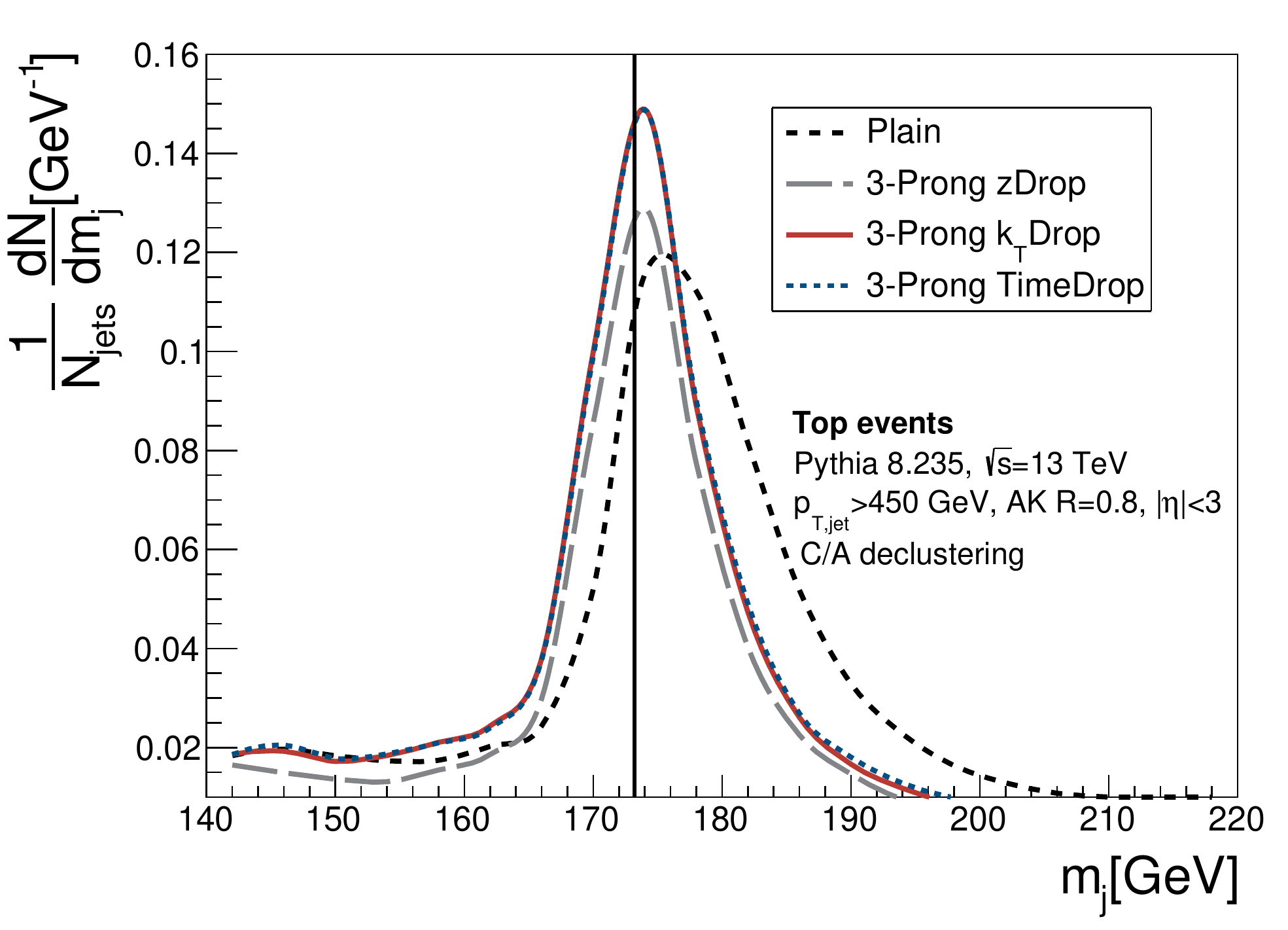}
\caption{The dynamically groomed mass distribution of top jets for $3$-Prong $z$Drop (dashed, gray), $k_T$Drop (solid, red) and TimeDrop (dotted, blue). For completeness, the vertical black line indicates the top quark mass and the plain distribution is given by the dashed, black line.}
\label{fig:mass-topdyg}
\end{figure}
\begin{figure*}
\includegraphics[scale=0.45]{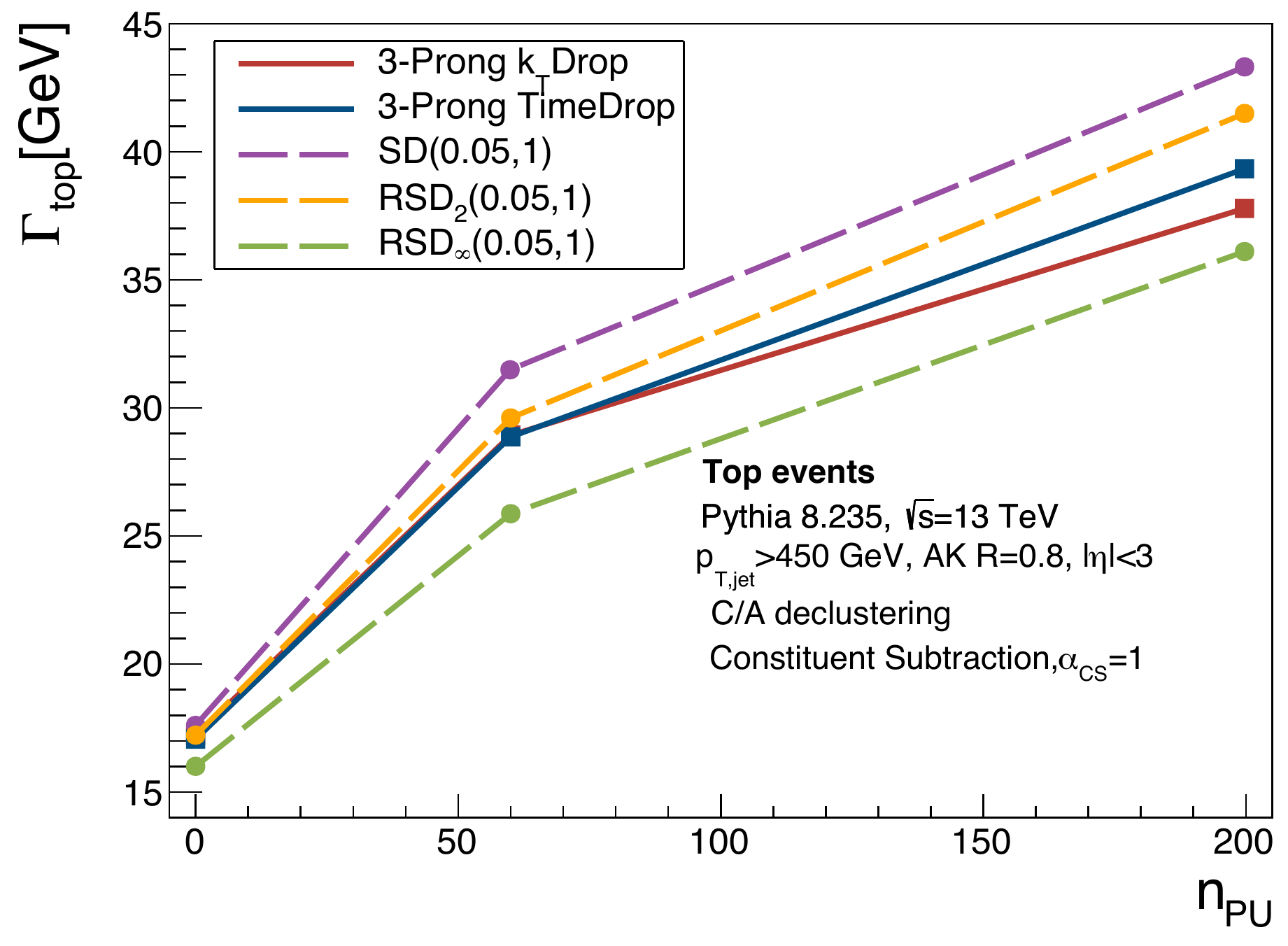}\quad\includegraphics[scale=0.45]{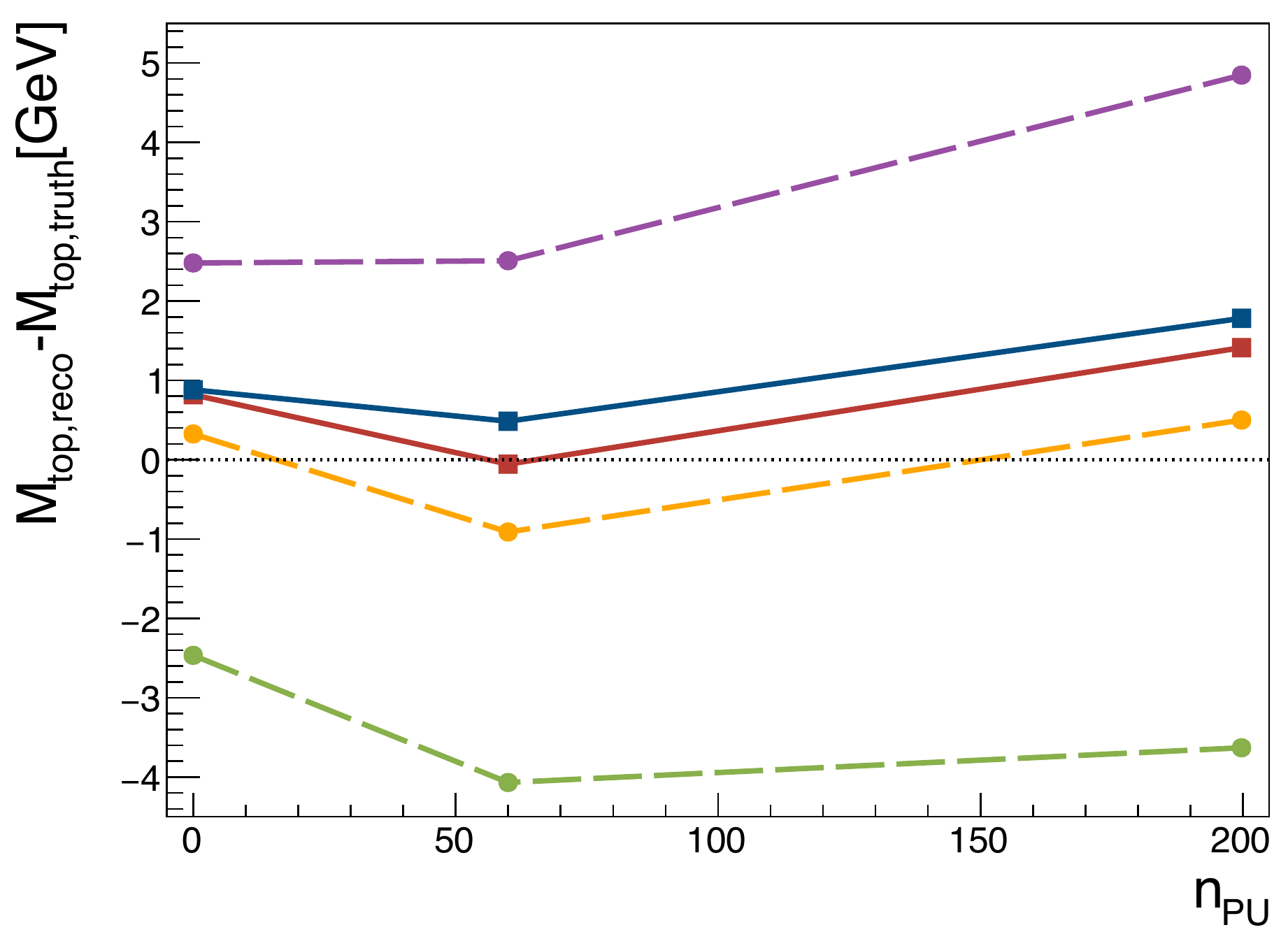}
\caption{Impact of pileup on the values of the top mass width (left) and its peak position with respect to the top quark mass $M_{{\rm{top,truth}}}$ (right) for different grooming methods at current LHC conditions ($n_{\rm_{PU}}\!=\!60$) and future HL-LHC ($n_{{\rm{PU}}}\!=\!200$).}
\label{fig:mass-top-comparison-wpu}
\end{figure*}
To study the performance of the 3-Prong version of dynamical grooming, we select the process $p\!+\!p\rightarrow t\bar{t}$. The jet mass spectrum in the top sample is shown in Fig.~\ref{fig:mass-topdyg}. The 3-Prong $k_T$Drop and TimeDrop curves are almost indistinguishable and give a reasonable description of the spectrum. In turn, 3-Prong $z$Drop results in a broader and shifted mass distribution. This fact together with the results for W jets described in the previous Section lead us to discard its use for non-QCD jets and refrain from showing its results in the rest of the manuscript.

To quantify the previous statements and gauge the role of pileup, we repeat the analysis described in Sec.~\ref{sec:w} and compute the peak position and the width of the top mass distribution. The results can be found on Fig.~\ref{fig:mass-top-comparison-wpu}. In the $n_\text{PU}\!=\!0$ case, $3$-Prong $k_T$Drop and TimeDrop remarkably give a more precise description of the top mass distribution than conventional SoftDrop. That is, for a similar value of the $\Gamma_{\rm{top}}$, the mass peak for $3$-Prong $k_T$Drop and TimeDrop is $1$~GeV closer to the resonance value than SD. We find particularly interesting to note that this would not have been the case with the default version of dynamical grooming. Hence, this result constitutes a solid support for the multi-prong strategy. Although with a different physics picture in mind, the addition of a second SoftDrop layer, RSD$_2$, also represents a neat improvement as we can clearly observe, in agreement with \cite{Dreyer:2018tjj}. However, when extending the number of layers to the fully recursive mode the resolution improvement comes at the cost of underestimating the mass peak. This trade-off could be potentially alleviated by a systematic study of the parameters ($z_{{\rm{cut}}},\beta$) in the SD condition, then introducing a Monte-Carlo dependence on experimental analyses. We would like to emphasize that this tuning exercise is completely needless in the dynamical grooming framework.

Regarding the impact of pileup contamination on the mass distribution of top events, we find that similarly to the previous Section, Run II pileup conditions ($n_\text{PU}\!=\!60$) are correctly handled by the two-step strategy of applying a particle-level background mitigation technique together with a grooming method such as 3-Prong $k_T$Drop, TimeDrop or RSD$_{2}$, whose results differ by less than $1$~GeV, in this particular case. Turning to the upcoming high-luminosity phase, it is worth emphasizing that the same three methods are successful in reproducing the mass peak position, although the distribution width is roughly doubled with respect to the ideal no pileup scenario. In particular,  3-Prong $k_T$Drop, TimeDrop values correspond to a slight improvement with respect to the RSD$_{2}$ case, suggesting an interesting resilience of the DyG family to pileup. Regarding RSD$_\infty$, we find a systematic underestimation of $M_{\rm{top}}$ that, from our point of view, does not offset the resolution improvement. We also observe a different trend on the evolution of $M_{\rm{top}}$ with increasing pileup in the top case with respect to the W (see Fig.~\ref{fig:mass-w-comparison-wpu}): while the W mass peak decreases with increasing pileup, $M_{\rm{top}}$ diminishes from $n_\text{PU}\!=\!0$ to $n_\text{PU}\!=\!60$ and then raises from $n_\text{PU}\!=\!60$ to $n_\text{PU}\!=\!200$. This is true for both SD and DyG, thus suggesting that it's as an effect caused by the background subtraction method rather than by the grooming technique. A promising tool to investigate this issue and further reduce the mass resolution on the pileup mitigation side could be the recently proposed event-wide version of Constituent Subtraction~\cite{Berta:2019hnj}.

\subsection{Boosted top tagging}
\label{sec:toptag}
\begin{figure*}
\includegraphics[scale=0.45]{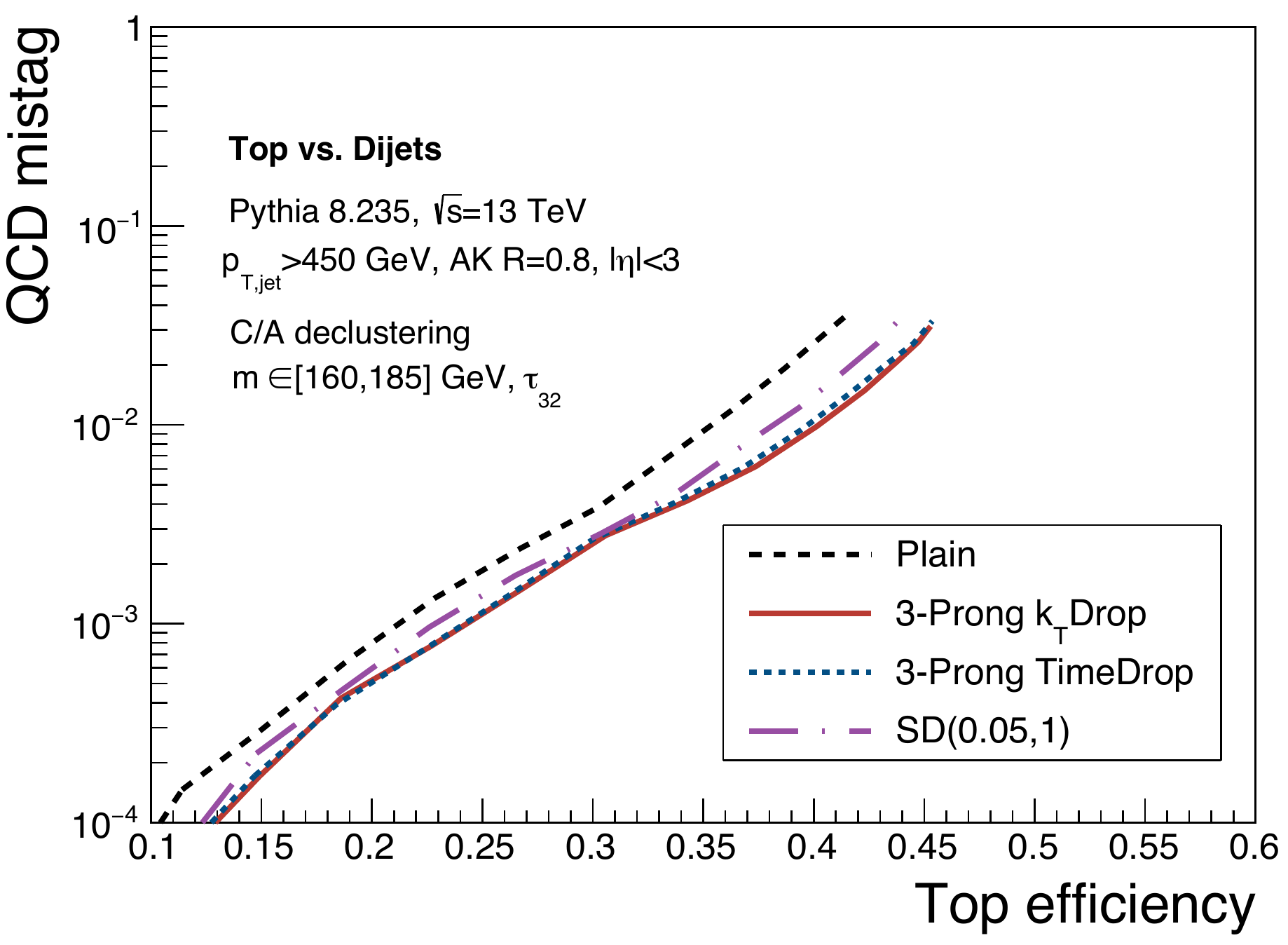}\includegraphics[scale=0.45]{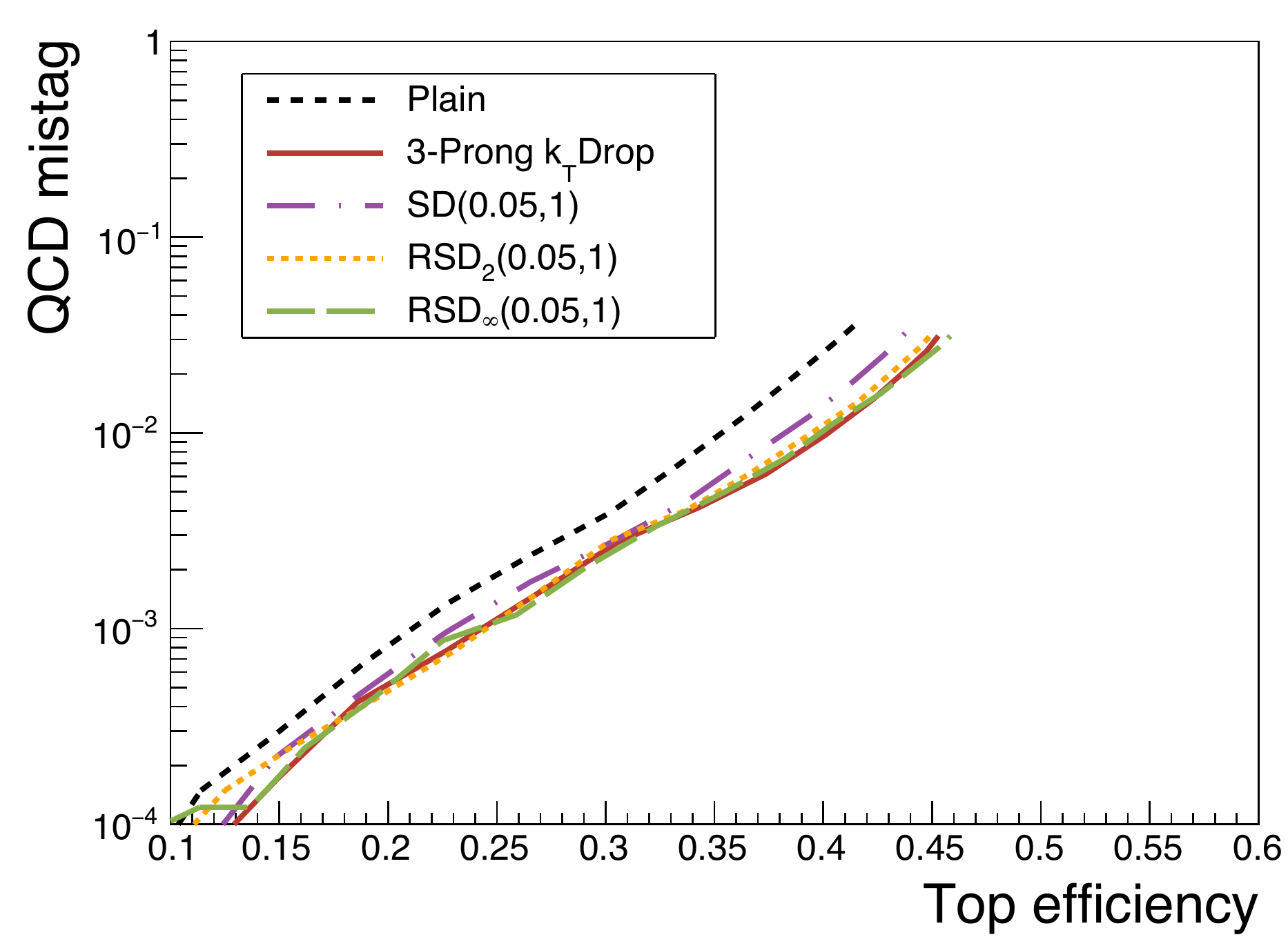}
\caption{Background mistag rate as a function of the signal efficiency (ROC curves) for the $3$-Prong version of Dynamical Grooming, SoftDrop and Recursive SoftDrop (described in Sec.~\ref{sec:grooming}).}
\label{fig:top-vs-dijets}
\end{figure*}
Finally, we study the tagging performance on the top sample. The main difference with respect to the analysis done in Sec.~\ref{sec:wtag} is that, following the line of reason of the N-subjettinnes discriminator, we choose $\tau_{32}$ as the jet shape variable in Eq.~\ref{eq:efficiency}.

On the left panel of Fig.~\ref{fig:top-vs-dijets}, the $3$-Prong dynamical grooming family is compared to SoftDrop. We find that $k_T$Drop gives the best signal-to-background ratio closely followed by TimeDrop, as suggested by Fig.~\ref{fig:mass-topdyg}. This result indicates the ability of the $3$-Prong DyG strategy to remove soft and wide angle radiation along the three prongs. Next, we compare the best performant of $3$-Prong DyG, $k_T$Drop, not only with SoftDrop but also with its recursive counterparts RSD$_2$ and RSD$_\infty$.  In this case, we observe a clear ordering in the results where RSD$_{\infty}$ clearly is the most efficient SD setting for this purpose, in quantitative agreement with the results obtained by~\cite{Dreyer:2018tjj}. Remarkably, $k_T$Drop, without the need of a fine-tuning procedure, results into the same top tagging performance as RSD family within the explored ($z_{{\rm cut}},\beta$)-values. The benefit of using a dynamical groomer to enhance the top tagging efficiency constitutes the main result of this paper and endorses $3$-Prong $k_T$Drop as a theoretically well grounded, yet easy to implement top tagger. 

\section{Discussion and Outlook}
\label{sec:conclusions}
This paper follows-up on a previous publication~\cite{Mehtar-Tani:2019rrk} where a novel dynamical grooming technique was proposed and applied to quark and gluon jets. In short, we select the hardest splitting on the primary Lund Plane of the C/A sequence and groom away all branches at larger angles. The one and only degree of freedom is the variable chosen to characterize the hardness: $z$, $k_T$ or $t_{\rm f}^{-1}$. In our first work, the aim was two-fold. First, to establish the analytical framework that allows to make predictions within perturbative QCD for this new class of substructure observables associated to the grooming technique. Then, to show that these observables, besides being analytically tractable, are also resilient to non-perturbative dynamics such as hadronization or underlying event, as was demonstrated through PYTHIA simulations. 

In this work we take a more phenomenological approach and explore the capability of dynamical grooming to distinguish W and top jets from QCD ones at LHC energies: an ubiquitous task at modern colliders. We compare the DyG results to the most popular grooming algorithm, namely SoftDrop. In addition, we consider the recently proposed recursive extension of SoftDrop that exhibited a better performance in boosted W/top tagging. Notice that the goal of this paper is not to asses (R)SD performance in an exhaustive way as a tuning of the parameters in the SD condition might lead to an improvement on the results shown in this paper, but rather use it as a reference point to baseline the performance of dynamical grooming. 

The main results of this manuscript together with some lines future work are described in what follows.
\begin{itemize}
\item While $2$-prong topologies can be handled by the out-of-the-box version of dynamical grooming, the characteristic $3$-prong decay of the top quark has lead us to introduce an extension dubbed $3$-prong DyG. In this case, not only the hardest but also the next-to-hardest splitting is considered. After finding which of the two happens at larger angles, all branches prior to it in the de-clustering sequence are groomed away plus the radiation in between the two. This procedure leads to a groomed jet built up of three subjets. The generalization of this algorithm to an arbitrary number of final state hard prongs could be implemented in a similar fashion and will be part of future extensions. Hence, the introduction of $3$-prong DyG represents the main conceptual development of this work.
\item One of the main goals of this paper is to identify which definition of hardness in Eq.~\eqref{eq:hadrness} gives a better performance, as it is the only parameter to be fixed in the context of DyG. The studies on W (top) mass reconstruction and tagging lead us to pin down ($3$-Prong) $k_T$Drop as the most robust and best performing setting of the DyG family. This adds up to its interesting analytical properties for QCD jets and the demonstrated resilience to hadronization, underlying event and pileup. In addition, a strong correlation between the hardest splitting in the C/A de-clustering sequence and the QCD shower at parton level has been recently shown. This might be partially responsible for the success of $k_T$Drop. Therefore, we strongly recommend the use of ($3$-Prong) $k_T$Drop in experimental analyses. We would like to emphasize that the lack of free parameters is absolutely relevant in scenarios where reliable Monte-Carlo simulations for tuning parameters are not possible such as model-independent BSM searches or heavy-ion collisions. 
\item The robustness of ($3$-Prong) $k_T$Drop against pileup, as shown in Figs.~\ref{fig:mass-w-comparison-wpu} and \ref{fig:mass-top-comparison-wpu}, together with its versatility, demonstrating a good performance on QCD, W and top jets, encourage us to consider it as a promising tool for studying jet modifications in hot and dense nuclear matter such as the one created in ultra-relativistic heavy-ion collisions, see e.g. \cite{Andrews:2018jcm}. In this environment, the reconstructed jet is embedded into an abundant thermal background that substantially contaminate precision measurements in the context of jet substructure. Hence, dynamical grooming could be used to i) efficiently remove soft and wide angle radiation and isolate the most perturbative splitting, ii) facilitate theory-to-data comparisons with DyG observables. These ideas will be explored in an upcoming publication. 
\end{itemize}

Finally, we plan to include both the dynamical groomer and its multi-prong version in the next release of {\tt{fastjet-contrib}}~\cite{fcontrib}. For the moment, the routines are publicly available at {\tt{https://github.com/aontoso/JetToyHI}} within the JetTools workshop setup~\cite{Andrews:2018jcm}. 
\section*{ACKNOWLEDGMENTS}
We are grateful to Frederic Dreyer for discussions on the Recursive SoftDrop method and to Jesse Thaler for a careful reading of the manuscript. The work of Y. M.-T. and A. S.-O. was supported by the U.S. Department of Energy, Office of Science, Office of Nuclear Physics, under contract No. DE- SC0012704,
and by Laboratory Directed Research and Development (LDRD) funds from Brookhaven Science Associates. Y. M.-T. acknowledges support from the RHIC Physics Fellow Program of the RIKEN BNL Research Center.
 K. T. is supported by a Starting Grant from Trond Mohn Foundation (BFS2018REK01) and the University of Bergen.

\bibliographystyle{apsrev4-1}
\bibliography{dyg_2}

\begin{thebibliography}{50}%
\makeatletter
\providecommand \@ifxundefined [1]{%
 \@ifx{#1\undefined}
}%
\providecommand \@ifnum [1]{%
 \ifnum #1\expandafter \@firstoftwo
 \else \expandafter \@secondoftwo
 \fi
}%
\providecommand \@ifx [1]{%
 \ifx #1\expandafter \@firstoftwo
 \else \expandafter \@secondoftwo
 \fi
}%
\providecommand \natexlab [1]{#1}%
\providecommand \enquote  [1]{``#1''}%
\providecommand \bibnamefont  [1]{#1}%
\providecommand \bibfnamefont [1]{#1}%
\providecommand \citenamefont [1]{#1}%
\providecommand \href@noop [0]{\@secondoftwo}%
\providecommand \href [0]{\begingroup \@sanitize@url \@href}%
\providecommand \@href[1]{\@@startlink{#1}\@@href}%
\providecommand \@@href[1]{\endgroup#1\@@endlink}%
\providecommand \@sanitize@url [0]{\catcode `\\12\catcode `\$12\catcode
  `\&12\catcode `\#12\catcode `\^12\catcode `\_12\catcode `\%12\relax}%
\providecommand \@@startlink[1]{}%
\providecommand \@@endlink[0]{}%
\providecommand \url  [0]{\begingroup\@sanitize@url \@url }%
\providecommand \@url [1]{\endgroup\@href {#1}{\urlprefix }}%
\providecommand \urlprefix  [0]{URL }%
\providecommand \Eprint [0]{\href }%
\providecommand \doibase [0]{http://dx.doi.org/}%
\providecommand \selectlanguage [0]{\@gobble}%
\providecommand \bibinfo  [0]{\@secondoftwo}%
\providecommand \bibfield  [0]{\@secondoftwo}%
\providecommand \translation [1]{[#1]}%
\providecommand \BibitemOpen [0]{}%
\providecommand \bibitemStop [0]{}%
\providecommand \bibitemNoStop [0]{.\EOS\space}%
\providecommand \EOS [0]{\spacefactor3000\relax}%
\providecommand \BibitemShut  [1]{\csname bibitem#1\endcsname}%
\let\auto@bib@innerbib\@empty
\bibitem [{\citenamefont {Mehtar-Tani}\ \emph {et~al.}(2020)\citenamefont
  {Mehtar-Tani}, \citenamefont {Soto-Ontoso},\ and\ \citenamefont
  {Tywoniuk}}]{Mehtar-Tani:2019rrk}%
  \BibitemOpen
  \bibfield  {author} {\bibinfo {author} {\bibfnamefont {Y.}~\bibnamefont
  {Mehtar-Tani}}, \bibinfo {author} {\bibfnamefont {A.}~\bibnamefont
  {Soto-Ontoso}}, \ and\ \bibinfo {author} {\bibfnamefont {K.}~\bibnamefont
  {Tywoniuk}},\ }\href {\doibase 10.1103/PhysRevD.101.034004} {\bibfield
  {journal} {\bibinfo  {journal} {Phys. Rev.}\ }\textbf {\bibinfo {volume}
  {D101}},\ \bibinfo {pages} {034004} (\bibinfo {year} {2020})},\ \Eprint
  {http://arxiv.org/abs/1911.00375} {arXiv:1911.00375 [hep-ph]} \BibitemShut
  {NoStop}%
\bibitem [{\citenamefont {Larkoski}\ \emph {et~al.}(2020)\citenamefont
  {Larkoski}, \citenamefont {Moult},\ and\ \citenamefont
  {Nachman}}]{Larkoski:2017jix}%
  \BibitemOpen
  \bibfield  {author} {\bibinfo {author} {\bibfnamefont {A.~J.}\ \bibnamefont
  {Larkoski}}, \bibinfo {author} {\bibfnamefont {I.}~\bibnamefont {Moult}}, \
  and\ \bibinfo {author} {\bibfnamefont {B.}~\bibnamefont {Nachman}},\ }\href
  {\doibase 10.1016/j.physrep.2019.11.001} {\bibfield  {journal} {\bibinfo
  {journal} {Phys. Rept.}\ }\textbf {\bibinfo {volume} {841}},\ \bibinfo
  {pages} {1} (\bibinfo {year} {2020})},\ \Eprint
  {http://arxiv.org/abs/1709.04464} {arXiv:1709.04464 [hep-ph]} \BibitemShut
  {NoStop}%
\bibitem [{\citenamefont {Marzani}\ \emph {et~al.}(2019)\citenamefont
  {Marzani}, \citenamefont {Soyez},\ and\ \citenamefont
  {Spannowsky}}]{Marzani:2019hun}%
  \BibitemOpen
  \bibfield  {author} {\bibinfo {author} {\bibfnamefont {S.}~\bibnamefont
  {Marzani}}, \bibinfo {author} {\bibfnamefont {G.}~\bibnamefont {Soyez}}, \
  and\ \bibinfo {author} {\bibfnamefont {M.}~\bibnamefont {Spannowsky}},\
  }\href {\doibase 10.1007/978-3-030-15709-8} {\  (\bibinfo {year} {2019}),\
  10.1007/978-3-030-15709-8},\ \bibinfo {note} {[Lect. Notes
  Phys.958,pp.(2019)]},\ \Eprint {http://arxiv.org/abs/1901.10342}
  {arXiv:1901.10342 [hep-ph]} \BibitemShut {NoStop}%
\bibitem [{Ben(2018)}]{Bendavid:2018nar}%
  \BibitemOpen
  \href@noop {} {\emph {\bibinfo {title} {{Les Houches 2017: Physics at TeV
  Colliders Standard Model Working Group Report}}}}\ (\bibinfo {year} {2018})\
  \Eprint {http://arxiv.org/abs/1803.07977} {arXiv:1803.07977 [hep-ph]}
  \BibitemShut {NoStop}%
\bibitem [{\citenamefont {Kogler}\ \emph {et~al.}(2019)\citenamefont {Kogler}
  \emph {et~al.}}]{Asquith:2018igt}%
  \BibitemOpen
  \bibfield  {author} {\bibinfo {author} {\bibfnamefont {R.}~\bibnamefont
  {Kogler}} \emph {et~al.},\ }\href {\doibase 10.1103/RevModPhys.91.045003}
  {\bibfield  {journal} {\bibinfo  {journal} {Rev. Mod. Phys.}\ }\textbf
  {\bibinfo {volume} {91}},\ \bibinfo {pages} {045003} (\bibinfo {year}
  {2019})},\ \Eprint {http://arxiv.org/abs/1803.06991} {arXiv:1803.06991
  [hep-ex]} \BibitemShut {NoStop}%
\bibitem [{\citenamefont {Kaplan}\ \emph {et~al.}(2008)\citenamefont {Kaplan},
  \citenamefont {Rehermann}, \citenamefont {Schwartz},\ and\ \citenamefont
  {Tweedie}}]{Kaplan:2008ie}%
  \BibitemOpen
  \bibfield  {author} {\bibinfo {author} {\bibfnamefont {D.~E.}\ \bibnamefont
  {Kaplan}}, \bibinfo {author} {\bibfnamefont {K.}~\bibnamefont {Rehermann}},
  \bibinfo {author} {\bibfnamefont {M.~D.}\ \bibnamefont {Schwartz}}, \ and\
  \bibinfo {author} {\bibfnamefont {B.}~\bibnamefont {Tweedie}},\ }\href
  {\doibase 10.1103/PhysRevLett.101.142001} {\bibfield  {journal} {\bibinfo
  {journal} {Phys.\ Rev.\ Lett.}\ }\textbf {\bibinfo {volume} {101}},\ \bibinfo
  {pages} {142001} (\bibinfo {year} {2008})},\ \Eprint
  {http://arxiv.org/abs/0806.0848} {arXiv:0806.0848 [hep-ph]} \BibitemShut
  {NoStop}%
\bibitem [{\citenamefont {Thaler}\ and\ \citenamefont
  {Van~Tilburg}(2012)}]{Thaler:2011gf}%
  \BibitemOpen
  \bibfield  {author} {\bibinfo {author} {\bibfnamefont {J.}~\bibnamefont
  {Thaler}}\ and\ \bibinfo {author} {\bibfnamefont {K.}~\bibnamefont
  {Van~Tilburg}},\ }\href {\doibase 10.1007/JHEP02(2012)093} {\bibfield
  {journal} {\bibinfo  {journal} {JHEP}\ }\textbf {\bibinfo {volume} {02}},\
  \bibinfo {pages} {093} (\bibinfo {year} {2012})},\ \Eprint
  {http://arxiv.org/abs/1108.2701} {arXiv:1108.2701 [hep-ph]} \BibitemShut
  {NoStop}%
\bibitem [{\citenamefont {Dasgupta}\ \emph
  {et~al.}(2013{\natexlab{a}})\citenamefont {Dasgupta}, \citenamefont
  {Fregoso}, \citenamefont {Marzani},\ and\ \citenamefont
  {Salam}}]{Dasgupta:2013ihk}%
  \BibitemOpen
  \bibfield  {author} {\bibinfo {author} {\bibfnamefont {M.}~\bibnamefont
  {Dasgupta}}, \bibinfo {author} {\bibfnamefont {A.}~\bibnamefont {Fregoso}},
  \bibinfo {author} {\bibfnamefont {S.}~\bibnamefont {Marzani}}, \ and\
  \bibinfo {author} {\bibfnamefont {G.~P.}\ \bibnamefont {Salam}},\ }\href
  {\doibase 10.1007/JHEP09(2013)029} {\bibfield  {journal} {\bibinfo  {journal}
  {JHEP}\ }\textbf {\bibinfo {volume} {09}},\ \bibinfo {pages} {029} (\bibinfo
  {year} {2013}{\natexlab{a}})},\ \Eprint {http://arxiv.org/abs/1307.0007}
  {arXiv:1307.0007 [hep-ph]} \BibitemShut {NoStop}%
\bibitem [{\citenamefont {Dasgupta}\ \emph
  {et~al.}(2013{\natexlab{b}})\citenamefont {Dasgupta}, \citenamefont
  {Fregoso}, \citenamefont {Marzani},\ and\ \citenamefont
  {Powling}}]{Dasgupta:2013via}%
  \BibitemOpen
  \bibfield  {author} {\bibinfo {author} {\bibfnamefont {M.}~\bibnamefont
  {Dasgupta}}, \bibinfo {author} {\bibfnamefont {A.}~\bibnamefont {Fregoso}},
  \bibinfo {author} {\bibfnamefont {S.}~\bibnamefont {Marzani}}, \ and\
  \bibinfo {author} {\bibfnamefont {A.}~\bibnamefont {Powling}},\ }\href
  {\doibase 10.1140/epjc/s10052-013-2623-3} {\bibfield  {journal} {\bibinfo
  {journal} {Eur. Phys. J.}\ }\textbf {\bibinfo {volume} {C73}},\ \bibinfo
  {pages} {2623} (\bibinfo {year} {2013}{\natexlab{b}})},\ \Eprint
  {http://arxiv.org/abs/1307.0013} {arXiv:1307.0013 [hep-ph]} \BibitemShut
  {NoStop}%
\bibitem [{\citenamefont {Napoletano}\ and\ \citenamefont
  {Soyez}(2018)}]{Napoletano:2018ohv}%
  \BibitemOpen
  \bibfield  {author} {\bibinfo {author} {\bibfnamefont {D.}~\bibnamefont
  {Napoletano}}\ and\ \bibinfo {author} {\bibfnamefont {G.}~\bibnamefont
  {Soyez}},\ }\href {\doibase 10.1007/JHEP12(2018)031} {\bibfield  {journal}
  {\bibinfo  {journal} {JHEP}\ }\textbf {\bibinfo {volume} {12}},\ \bibinfo
  {pages} {031} (\bibinfo {year} {2018})},\ \Eprint
  {http://arxiv.org/abs/1809.04602} {arXiv:1809.04602 [hep-ph]} \BibitemShut
  {NoStop}%
\bibitem [{\citenamefont {Butter}\ \emph {et~al.}(2019)\citenamefont {Butter}
  \emph {et~al.}}]{Kasieczka:2019dbj}%
  \BibitemOpen
  \bibfield  {author} {\bibinfo {author} {\bibfnamefont {A.}~\bibnamefont
  {Butter}} \emph {et~al.},\ }\href {\doibase 10.21468/SciPostPhys.7.1.014}
  {\bibfield  {journal} {\bibinfo  {journal} {SciPost Phys.}\ }\textbf
  {\bibinfo {volume} {7}},\ \bibinfo {pages} {014} (\bibinfo {year} {2019})},\
  \Eprint {http://arxiv.org/abs/1902.09914} {arXiv:1902.09914 [hep-ph]}
  \BibitemShut {NoStop}%
\bibitem [{\citenamefont {Collins}\ \emph {et~al.}(2019)\citenamefont
  {Collins}, \citenamefont {Howe},\ and\ \citenamefont
  {Nachman}}]{Collins:2019jip}%
  \BibitemOpen
  \bibfield  {author} {\bibinfo {author} {\bibfnamefont {J.~H.}\ \bibnamefont
  {Collins}}, \bibinfo {author} {\bibfnamefont {K.}~\bibnamefont {Howe}}, \
  and\ \bibinfo {author} {\bibfnamefont {B.}~\bibnamefont {Nachman}},\ }\href
  {\doibase 10.1103/PhysRevD.99.014038} {\bibfield  {journal} {\bibinfo
  {journal} {Phys.\ Rev.\ D}\ }\textbf {\bibinfo {volume} {99}},\ \bibinfo
  {pages} {014038} (\bibinfo {year} {2019})},\ \Eprint
  {http://arxiv.org/abs/1902.02634} {arXiv:1902.02634 [hep-ph]} \BibitemShut
  {NoStop}%
\bibitem [{\citenamefont {Andreassen}\ \emph {et~al.}(2020)\citenamefont
  {Andreassen}, \citenamefont {Nachman},\ and\ \citenamefont
  {Shih}}]{Andreassen:2020nkr}%
  \BibitemOpen
  \bibfield  {author} {\bibinfo {author} {\bibfnamefont {A.}~\bibnamefont
  {Andreassen}}, \bibinfo {author} {\bibfnamefont {B.}~\bibnamefont {Nachman}},
  \ and\ \bibinfo {author} {\bibfnamefont {D.}~\bibnamefont {Shih}},\
  }\href@noop {} {\  (\bibinfo {year} {2020})},\ \Eprint
  {http://arxiv.org/abs/2001.05001} {arXiv:2001.05001 [hep-ph]} \BibitemShut
  {NoStop}%
\bibitem [{\citenamefont {Nachman}\ and\ \citenamefont
  {Shih}(2020)}]{Nachman:2020lpy}%
  \BibitemOpen
  \bibfield  {author} {\bibinfo {author} {\bibfnamefont {B.}~\bibnamefont
  {Nachman}}\ and\ \bibinfo {author} {\bibfnamefont {D.}~\bibnamefont {Shih}},\
  }\href {\doibase 10.1103/PhysRevD.101.075042} {\bibfield  {journal} {\bibinfo
   {journal} {Phys. Rev. D}\ }\textbf {\bibinfo {volume} {101}},\ \bibinfo
  {pages} {075042} (\bibinfo {year} {2020})},\ \Eprint
  {http://arxiv.org/abs/2001.04990} {arXiv:2001.04990 [hep-ph]} \BibitemShut
  {NoStop}%
\bibitem [{\citenamefont {Chien}\ and\ \citenamefont
  {Stewart}(2019)}]{Chien:2019osu}%
  \BibitemOpen
  \bibfield  {author} {\bibinfo {author} {\bibfnamefont {Y.-T.}\ \bibnamefont
  {Chien}}\ and\ \bibinfo {author} {\bibfnamefont {I.~W.}\ \bibnamefont
  {Stewart}},\ }\href@noop {} {\  (\bibinfo {year} {2019})},\ \Eprint
  {http://arxiv.org/abs/1907.11107} {arXiv:1907.11107 [hep-ph]} \BibitemShut
  {NoStop}%
\bibitem [{\citenamefont {Carrazza}\ and\ \citenamefont
  {Dreyer}(2019)}]{Carrazza:2019efs}%
  \BibitemOpen
  \bibfield  {author} {\bibinfo {author} {\bibfnamefont {S.}~\bibnamefont
  {Carrazza}}\ and\ \bibinfo {author} {\bibfnamefont {F.~A.}\ \bibnamefont
  {Dreyer}},\ }\href {\doibase 10.1103/PhysRevD.100.014014} {\bibfield
  {journal} {\bibinfo  {journal} {Phys. Rev.}\ }\textbf {\bibinfo {volume}
  {D100}},\ \bibinfo {pages} {014014} (\bibinfo {year} {2019})},\ \Eprint
  {http://arxiv.org/abs/1903.09644} {arXiv:1903.09644 [hep-ph]} \BibitemShut
  {NoStop}%
\bibitem [{\citenamefont {Dreyer}\ \emph
  {et~al.}(2018{\natexlab{a}})\citenamefont {Dreyer}, \citenamefont {Necib},
  \citenamefont {Soyez},\ and\ \citenamefont {Thaler}}]{Dreyer:2018tjj}%
  \BibitemOpen
  \bibfield  {author} {\bibinfo {author} {\bibfnamefont {F.~A.}\ \bibnamefont
  {Dreyer}}, \bibinfo {author} {\bibfnamefont {L.}~\bibnamefont {Necib}},
  \bibinfo {author} {\bibfnamefont {G.}~\bibnamefont {Soyez}}, \ and\ \bibinfo
  {author} {\bibfnamefont {J.}~\bibnamefont {Thaler}},\ }\href {\doibase
  10.1007/JHEP06(2018)093} {\bibfield  {journal} {\bibinfo  {journal} {JHEP}\
  }\textbf {\bibinfo {volume} {06}},\ \bibinfo {pages} {093} (\bibinfo {year}
  {2018}{\natexlab{a}})},\ \Eprint {http://arxiv.org/abs/1804.03657}
  {arXiv:1804.03657 [hep-ph]} \BibitemShut {NoStop}%
\bibitem [{\citenamefont {Larkoski}\ \emph {et~al.}(2014)\citenamefont
  {Larkoski}, \citenamefont {Marzani}, \citenamefont {Soyez},\ and\
  \citenamefont {Thaler}}]{Larkoski:2014wba}%
  \BibitemOpen
  \bibfield  {author} {\bibinfo {author} {\bibfnamefont {A.~J.}\ \bibnamefont
  {Larkoski}}, \bibinfo {author} {\bibfnamefont {S.}~\bibnamefont {Marzani}},
  \bibinfo {author} {\bibfnamefont {G.}~\bibnamefont {Soyez}}, \ and\ \bibinfo
  {author} {\bibfnamefont {J.}~\bibnamefont {Thaler}},\ }\href {\doibase
  10.1007/JHEP05(2014)146} {\bibfield  {journal} {\bibinfo  {journal} {JHEP}\
  }\textbf {\bibinfo {volume} {05}},\ \bibinfo {pages} {146} (\bibinfo {year}
  {2014})},\ \Eprint {http://arxiv.org/abs/1402.2657} {arXiv:1402.2657
  [hep-ph]} \BibitemShut {NoStop}%
\bibitem [{\citenamefont {Butterworth}\ \emph {et~al.}(2008)\citenamefont
  {Butterworth}, \citenamefont {Davison}, \citenamefont {Rubin},\ and\
  \citenamefont {Salam}}]{Butterworth:2008iy}%
  \BibitemOpen
  \bibfield  {author} {\bibinfo {author} {\bibfnamefont {J.~M.}\ \bibnamefont
  {Butterworth}}, \bibinfo {author} {\bibfnamefont {A.~R.}\ \bibnamefont
  {Davison}}, \bibinfo {author} {\bibfnamefont {M.}~\bibnamefont {Rubin}}, \
  and\ \bibinfo {author} {\bibfnamefont {G.~P.}\ \bibnamefont {Salam}},\ }\href
  {\doibase 10.1103/PhysRevLett.100.242001} {\bibfield  {journal} {\bibinfo
  {journal} {Phys. Rev. Lett.}\ }\textbf {\bibinfo {volume} {100}},\ \bibinfo
  {pages} {242001} (\bibinfo {year} {2008})},\ \Eprint
  {http://arxiv.org/abs/0802.2470} {arXiv:0802.2470 [hep-ph]} \BibitemShut
  {NoStop}%
\bibitem [{\citenamefont {Kang}\ \emph {et~al.}(2018)\citenamefont {Kang},
  \citenamefont {Lee}, \citenamefont {Liu},\ and\ \citenamefont
  {Ringer}}]{Kang:2018jwa}%
  \BibitemOpen
  \bibfield  {author} {\bibinfo {author} {\bibfnamefont {Z.-B.}\ \bibnamefont
  {Kang}}, \bibinfo {author} {\bibfnamefont {K.}~\bibnamefont {Lee}}, \bibinfo
  {author} {\bibfnamefont {X.}~\bibnamefont {Liu}}, \ and\ \bibinfo {author}
  {\bibfnamefont {F.}~\bibnamefont {Ringer}},\ }\href {\doibase
  10.1007/JHEP10(2018)137} {\bibfield  {journal} {\bibinfo  {journal} {JHEP}\
  }\textbf {\bibinfo {volume} {10}},\ \bibinfo {pages} {137} (\bibinfo {year}
  {2018})},\ \Eprint {http://arxiv.org/abs/1803.03645} {arXiv:1803.03645
  [hep-ph]} \BibitemShut {NoStop}%
\bibitem [{\citenamefont {Marzani}\ \emph {et~al.}(2018)\citenamefont
  {Marzani}, \citenamefont {Schunk},\ and\ \citenamefont
  {Soyez}}]{Marzani:2017kqd}%
  \BibitemOpen
  \bibfield  {author} {\bibinfo {author} {\bibfnamefont {S.}~\bibnamefont
  {Marzani}}, \bibinfo {author} {\bibfnamefont {L.}~\bibnamefont {Schunk}}, \
  and\ \bibinfo {author} {\bibfnamefont {G.}~\bibnamefont {Soyez}},\ }\href
  {\doibase 10.1140/epjc/s10052-018-5579-5} {\bibfield  {journal} {\bibinfo
  {journal} {Eur. Phys. J.}\ }\textbf {\bibinfo {volume} {C78}},\ \bibinfo
  {pages} {96} (\bibinfo {year} {2018})},\ \Eprint
  {http://arxiv.org/abs/1712.05105} {arXiv:1712.05105 [hep-ph]} \BibitemShut
  {NoStop}%
\bibitem [{\citenamefont {Ellis}\ \emph {et~al.}(2010)\citenamefont {Ellis},
  \citenamefont {Vermilion},\ and\ \citenamefont {Walsh}}]{PhysRevD.81.094023}%
  \BibitemOpen
  \bibfield  {author} {\bibinfo {author} {\bibfnamefont {S.~D.}\ \bibnamefont
  {Ellis}}, \bibinfo {author} {\bibfnamefont {C.~K.}\ \bibnamefont
  {Vermilion}}, \ and\ \bibinfo {author} {\bibfnamefont {J.~R.}\ \bibnamefont
  {Walsh}},\ }\href {\doibase 10.1103/PhysRevD.81.094023} {\bibfield  {journal}
  {\bibinfo  {journal} {Phys. Rev. D}\ }\textbf {\bibinfo {volume} {81}},\
  \bibinfo {pages} {094023} (\bibinfo {year} {2010})}\BibitemShut {NoStop}%
\bibitem [{\citenamefont {Krohn}\ \emph {et~al.}(2010)\citenamefont {Krohn},
  \citenamefont {Thaler},\ and\ \citenamefont {Wang}}]{Krohn:2009th}%
  \BibitemOpen
  \bibfield  {author} {\bibinfo {author} {\bibfnamefont {D.}~\bibnamefont
  {Krohn}}, \bibinfo {author} {\bibfnamefont {J.}~\bibnamefont {Thaler}}, \
  and\ \bibinfo {author} {\bibfnamefont {L.-T.}\ \bibnamefont {Wang}},\ }\href
  {\doibase 10.1007/JHEP02(2010)084} {\bibfield  {journal} {\bibinfo  {journal}
  {JHEP}\ }\textbf {\bibinfo {volume} {02}},\ \bibinfo {pages} {084} (\bibinfo
  {year} {2010})},\ \Eprint {http://arxiv.org/abs/0912.1342} {arXiv:0912.1342
  [hep-ph]} \BibitemShut {NoStop}%
\bibitem [{\citenamefont {Thaler}\ and\ \citenamefont
  {Van~Tilburg}(2011)}]{Thaler:2010tr}%
  \BibitemOpen
  \bibfield  {author} {\bibinfo {author} {\bibfnamefont {J.}~\bibnamefont
  {Thaler}}\ and\ \bibinfo {author} {\bibfnamefont {K.}~\bibnamefont
  {Van~Tilburg}},\ }\href {\doibase 10.1007/JHEP03(2011)015} {\bibfield
  {journal} {\bibinfo  {journal} {JHEP}\ }\textbf {\bibinfo {volume} {03}},\
  \bibinfo {pages} {015} (\bibinfo {year} {2011})},\ \Eprint
  {http://arxiv.org/abs/1011.2268} {arXiv:1011.2268 [hep-ph]} \BibitemShut
  {NoStop}%
\bibitem [{\citenamefont {Larkoski}\ \emph {et~al.}(2013)\citenamefont
  {Larkoski}, \citenamefont {Salam},\ and\ \citenamefont
  {Thaler}}]{Larkoski:2013eya}%
  \BibitemOpen
  \bibfield  {author} {\bibinfo {author} {\bibfnamefont {A.~J.}\ \bibnamefont
  {Larkoski}}, \bibinfo {author} {\bibfnamefont {G.~P.}\ \bibnamefont {Salam}},
  \ and\ \bibinfo {author} {\bibfnamefont {J.}~\bibnamefont {Thaler}},\ }\href
  {\doibase 10.1007/JHEP06(2013)108} {\bibfield  {journal} {\bibinfo  {journal}
  {JHEP}\ }\textbf {\bibinfo {volume} {06}},\ \bibinfo {pages} {108} (\bibinfo
  {year} {2013})},\ \Eprint {http://arxiv.org/abs/1305.0007} {arXiv:1305.0007
  [hep-ph]} \BibitemShut {NoStop}%
\bibitem [{\citenamefont {Komiske}\ \emph {et~al.}(2018)\citenamefont
  {Komiske}, \citenamefont {Metodiev},\ and\ \citenamefont
  {Thaler}}]{Komiske:2017aww}%
  \BibitemOpen
  \bibfield  {author} {\bibinfo {author} {\bibfnamefont {P.~T.}\ \bibnamefont
  {Komiske}}, \bibinfo {author} {\bibfnamefont {E.~M.}\ \bibnamefont
  {Metodiev}}, \ and\ \bibinfo {author} {\bibfnamefont {J.}~\bibnamefont
  {Thaler}},\ }\href {\doibase 10.1007/JHEP04(2018)013} {\bibfield  {journal}
  {\bibinfo  {journal} {JHEP}\ }\textbf {\bibinfo {volume} {04}},\ \bibinfo
  {pages} {013} (\bibinfo {year} {2018})},\ \Eprint
  {http://arxiv.org/abs/1712.07124} {arXiv:1712.07124 [hep-ph]} \BibitemShut
  {NoStop}%
\bibitem [{\citenamefont {Aad}\ \emph {et~al.}(2016)\citenamefont {Aad} \emph
  {et~al.}}]{Aad:2015rpa}%
  \BibitemOpen
  \bibfield  {author} {\bibinfo {author} {\bibfnamefont {G.}~\bibnamefont
  {Aad}} \emph {et~al.} (\bibinfo {collaboration} {ATLAS}),\ }\href {\doibase
  10.1140/epjc/s10052-016-3978-z} {\bibfield  {journal} {\bibinfo  {journal}
  {Eur. Phys. J.}\ }\textbf {\bibinfo {volume} {C76}},\ \bibinfo {pages} {154}
  (\bibinfo {year} {2016})},\ \Eprint {http://arxiv.org/abs/1510.05821}
  {arXiv:1510.05821 [hep-ex]} \BibitemShut {NoStop}%
\bibitem [{\citenamefont {Sirunyan}\ \emph {et~al.}(2020)\citenamefont
  {Sirunyan} \emph {et~al.}}]{Sirunyan:2019qia}%
  \BibitemOpen
  \bibfield  {author} {\bibinfo {author} {\bibfnamefont {A.~M.}\ \bibnamefont
  {Sirunyan}} \emph {et~al.} (\bibinfo {collaboration} {CMS}),\ }\href
  {\doibase 10.1007/JHEP03(2020)131} {\bibfield  {journal} {\bibinfo  {journal}
  {JHEP}\ }\textbf {\bibinfo {volume} {03}},\ \bibinfo {pages} {131} (\bibinfo
  {year} {2020})},\ \Eprint {http://arxiv.org/abs/1912.01662} {arXiv:1912.01662
  [hep-ex]} \BibitemShut {NoStop}%
\bibitem [{\citenamefont {Schramm}(2018)}]{Schramm:2018uyb}%
  \BibitemOpen
  \bibfield  {author} {\bibinfo {author} {\bibfnamefont {S.}~\bibnamefont
  {Schramm}} (\bibinfo {collaboration} {ATLAS, CMS}),\ }\href {\doibase
  10.22323/1.336.0192} {\bibfield  {journal} {\bibinfo  {journal} {PoS}\
  }\textbf {\bibinfo {volume} {Confinement2018}},\ \bibinfo {pages} {192}
  (\bibinfo {year} {2018})}\BibitemShut {NoStop}%
\bibitem [{\citenamefont {Aad}\ \emph {et~al.}(2020)\citenamefont {Aad} \emph
  {et~al.}}]{Aad:2019vyi}%
  \BibitemOpen
  \bibfield  {author} {\bibinfo {author} {\bibfnamefont {G.}~\bibnamefont
  {Aad}} \emph {et~al.} (\bibinfo {collaboration} {ATLAS}),\ }\href {\doibase
  10.1103/PhysRevD.101.052007} {\bibfield  {journal} {\bibinfo  {journal}
  {Phys. Rev. D}\ }\textbf {\bibinfo {volume} {101}},\ \bibinfo {pages}
  {052007} (\bibinfo {year} {2020})},\ \Eprint
  {http://arxiv.org/abs/1912.09837} {arXiv:1912.09837 [hep-ex]} \BibitemShut
  {NoStop}%
\bibitem [{\citenamefont {Adam}\ \emph {et~al.}(2020)\citenamefont {Adam} \emph
  {et~al.}}]{Adam:2020kug}%
  \BibitemOpen
  \bibfield  {author} {\bibinfo {author} {\bibfnamefont {J.}~\bibnamefont
  {Adam}} \emph {et~al.} (\bibinfo {collaboration} {STAR}),\ }\href@noop {} {\
  (\bibinfo {year} {2020})},\ \Eprint {http://arxiv.org/abs/2003.02114}
  {arXiv:2003.02114 [hep-ex]} \BibitemShut {NoStop}%
\bibitem [{\citenamefont {Kang}\ \emph {et~al.}(2020)\citenamefont {Kang},
  \citenamefont {Lee}, \citenamefont {Liu}, \citenamefont {Neill},\ and\
  \citenamefont {Ringer}}]{Kang:2019prh}%
  \BibitemOpen
  \bibfield  {author} {\bibinfo {author} {\bibfnamefont {Z.-B.}\ \bibnamefont
  {Kang}}, \bibinfo {author} {\bibfnamefont {K.}~\bibnamefont {Lee}}, \bibinfo
  {author} {\bibfnamefont {X.}~\bibnamefont {Liu}}, \bibinfo {author}
  {\bibfnamefont {D.}~\bibnamefont {Neill}}, \ and\ \bibinfo {author}
  {\bibfnamefont {F.}~\bibnamefont {Ringer}},\ }\href {\doibase
  10.1007/JHEP02(2020)054} {\bibfield  {journal} {\bibinfo  {journal} {JHEP}\
  }\textbf {\bibinfo {volume} {02}},\ \bibinfo {pages} {054} (\bibinfo {year}
  {2020})},\ \Eprint {http://arxiv.org/abs/1908.01783} {arXiv:1908.01783
  [hep-ph]} \BibitemShut {NoStop}%
\bibitem [{\citenamefont {Dokshitzer}\ \emph {et~al.}(1997)\citenamefont
  {Dokshitzer}, \citenamefont {Leder}, \citenamefont {Moretti},\ and\
  \citenamefont {Webber}}]{Dokshitzer:1997in}%
  \BibitemOpen
  \bibfield  {author} {\bibinfo {author} {\bibfnamefont {Y.~L.}\ \bibnamefont
  {Dokshitzer}}, \bibinfo {author} {\bibfnamefont {G.~D.}\ \bibnamefont
  {Leder}}, \bibinfo {author} {\bibfnamefont {S.}~\bibnamefont {Moretti}}, \
  and\ \bibinfo {author} {\bibfnamefont {B.~R.}\ \bibnamefont {Webber}},\
  }\href {\doibase 10.1088/1126-6708/1997/08/001} {\bibfield  {journal}
  {\bibinfo  {journal} {JHEP}\ }\textbf {\bibinfo {volume} {08}},\ \bibinfo
  {pages} {001} (\bibinfo {year} {1997})},\ \Eprint
  {http://arxiv.org/abs/hep-ph/9707323} {arXiv:hep-ph/9707323 [hep-ph]}
  \BibitemShut {NoStop}%
\bibitem [{\citenamefont {Frye}\ \emph {et~al.}(2017)\citenamefont {Frye},
  \citenamefont {Larkoski}, \citenamefont {Thaler},\ and\ \citenamefont
  {Zhou}}]{Frye:2017yrw}%
  \BibitemOpen
  \bibfield  {author} {\bibinfo {author} {\bibfnamefont {C.}~\bibnamefont
  {Frye}}, \bibinfo {author} {\bibfnamefont {A.~J.}\ \bibnamefont {Larkoski}},
  \bibinfo {author} {\bibfnamefont {J.}~\bibnamefont {Thaler}}, \ and\ \bibinfo
  {author} {\bibfnamefont {K.}~\bibnamefont {Zhou}},\ }\href {\doibase
  10.1007/JHEP09(2017)083} {\bibfield  {journal} {\bibinfo  {journal} {JHEP}\
  }\textbf {\bibinfo {volume} {09}},\ \bibinfo {pages} {083} (\bibinfo {year}
  {2017})},\ \Eprint {http://arxiv.org/abs/1704.06266} {arXiv:1704.06266
  [hep-ph]} \BibitemShut {NoStop}%
\bibitem [{\citenamefont {Acharya}\ \emph {et~al.}(2020)\citenamefont {Acharya}
  \emph {et~al.}}]{Acharya:2019djg}%
  \BibitemOpen
  \bibfield  {author} {\bibinfo {author} {\bibfnamefont {S.}~\bibnamefont
  {Acharya}} \emph {et~al.} (\bibinfo {collaboration} {ALICE}),\ }\href
  {\doibase 10.1016/j.physletb.2020.135227} {\bibfield  {journal} {\bibinfo
  {journal} {Phys. Lett. B}\ }\textbf {\bibinfo {volume} {802}},\ \bibinfo
  {pages} {135227} (\bibinfo {year} {2020})},\ \Eprint
  {http://arxiv.org/abs/1905.02512} {arXiv:1905.02512 [nucl-ex]} \BibitemShut
  {NoStop}%
\bibitem [{\citenamefont {Casalderrey-Solana}\ \emph
  {et~al.}(2020)\citenamefont {Casalderrey-Solana}, \citenamefont {Milhano},
  \citenamefont {Pablos},\ and\ \citenamefont
  {Rajagopal}}]{Casalderrey-Solana:2019ubu}%
  \BibitemOpen
  \bibfield  {author} {\bibinfo {author} {\bibfnamefont {J.}~\bibnamefont
  {Casalderrey-Solana}}, \bibinfo {author} {\bibfnamefont {G.}~\bibnamefont
  {Milhano}}, \bibinfo {author} {\bibfnamefont {D.}~\bibnamefont {Pablos}}, \
  and\ \bibinfo {author} {\bibfnamefont {K.}~\bibnamefont {Rajagopal}},\ }\href
  {\doibase 10.1007/JHEP01(2020)044} {\bibfield  {journal} {\bibinfo  {journal}
  {JHEP}\ }\textbf {\bibinfo {volume} {01}},\ \bibinfo {pages} {044} (\bibinfo
  {year} {2020})},\ \Eprint {http://arxiv.org/abs/1907.11248} {arXiv:1907.11248
  [hep-ph]} \BibitemShut {NoStop}%
\bibitem [{\citenamefont {Caucal}\ \emph {et~al.}(2019)\citenamefont {Caucal},
  \citenamefont {Iancu},\ and\ \citenamefont {Soyez}}]{Caucal:2019uvr}%
  \BibitemOpen
  \bibfield  {author} {\bibinfo {author} {\bibfnamefont {P.}~\bibnamefont
  {Caucal}}, \bibinfo {author} {\bibfnamefont {E.}~\bibnamefont {Iancu}}, \
  and\ \bibinfo {author} {\bibfnamefont {G.}~\bibnamefont {Soyez}},\ }\href
  {\doibase 10.1007/JHEP10(2019)273} {\bibfield  {journal} {\bibinfo  {journal}
  {JHEP}\ }\textbf {\bibinfo {volume} {10}},\ \bibinfo {pages} {273} (\bibinfo
  {year} {2019})},\ \Eprint {http://arxiv.org/abs/1907.04866} {arXiv:1907.04866
  [hep-ph]} \BibitemShut {NoStop}%
\bibitem [{\citenamefont {Andersson}\ \emph {et~al.}(1989)\citenamefont
  {Andersson}, \citenamefont {Gustafson}, \citenamefont {Lonnblad},\ and\
  \citenamefont {Pettersson}}]{Andersson:1988gp}%
  \BibitemOpen
  \bibfield  {author} {\bibinfo {author} {\bibfnamefont {B.}~\bibnamefont
  {Andersson}}, \bibinfo {author} {\bibfnamefont {G.}~\bibnamefont
  {Gustafson}}, \bibinfo {author} {\bibfnamefont {L.}~\bibnamefont {Lonnblad}},
  \ and\ \bibinfo {author} {\bibfnamefont {U.}~\bibnamefont {Pettersson}},\
  }\href {\doibase 10.1007/BF01550942} {\bibfield  {journal} {\bibinfo
  {journal} {Z. Phys.}\ }\textbf {\bibinfo {volume} {C43}},\ \bibinfo {pages}
  {625} (\bibinfo {year} {1989})}\BibitemShut {NoStop}%
\bibitem [{\citenamefont {Dreyer}\ \emph
  {et~al.}(2018{\natexlab{b}})\citenamefont {Dreyer}, \citenamefont {Salam},\
  and\ \citenamefont {Soyez}}]{Dreyer:2018nbf}%
  \BibitemOpen
  \bibfield  {author} {\bibinfo {author} {\bibfnamefont {F.~A.}\ \bibnamefont
  {Dreyer}}, \bibinfo {author} {\bibfnamefont {G.~P.}\ \bibnamefont {Salam}}, \
  and\ \bibinfo {author} {\bibfnamefont {G.}~\bibnamefont {Soyez}},\ }\href
  {\doibase 10.1007/JHEP12(2018)064} {\bibfield  {journal} {\bibinfo  {journal}
  {JHEP}\ }\textbf {\bibinfo {volume} {12}},\ \bibinfo {pages} {064} (\bibinfo
  {year} {2018}{\natexlab{b}})},\ \Eprint {http://arxiv.org/abs/1807.04758}
  {arXiv:1807.04758 [hep-ph]} \BibitemShut {NoStop}%
\bibitem [{\citenamefont {Sjostrand}\ \emph {et~al.}(2008)\citenamefont
  {Sjostrand}, \citenamefont {Mrenna},\ and\ \citenamefont
  {Skands}}]{Sjostrand:2007gs}%
  \BibitemOpen
  \bibfield  {author} {\bibinfo {author} {\bibfnamefont {T.}~\bibnamefont
  {Sjostrand}}, \bibinfo {author} {\bibfnamefont {S.}~\bibnamefont {Mrenna}}, \
  and\ \bibinfo {author} {\bibfnamefont {P.~Z.}\ \bibnamefont {Skands}},\
  }\href {\doibase 10.1016/j.cpc.2008.01.036} {\bibfield  {journal} {\bibinfo
  {journal} {Comput. Phys. Commun.}\ }\textbf {\bibinfo {volume} {178}},\
  \bibinfo {pages} {852} (\bibinfo {year} {2008})},\ \Eprint
  {http://arxiv.org/abs/0710.3820} {arXiv:0710.3820 [hep-ph]} \BibitemShut
  {NoStop}%
\bibitem [{\citenamefont {Cacciari}\ \emph {et~al.}(2008)\citenamefont
  {Cacciari}, \citenamefont {Salam},\ and\ \citenamefont
  {Soyez}}]{Cacciari:2008gp}%
  \BibitemOpen
  \bibfield  {author} {\bibinfo {author} {\bibfnamefont {M.}~\bibnamefont
  {Cacciari}}, \bibinfo {author} {\bibfnamefont {G.~P.}\ \bibnamefont {Salam}},
  \ and\ \bibinfo {author} {\bibfnamefont {G.}~\bibnamefont {Soyez}},\ }\href
  {\doibase 10.1088/1126-6708/2008/04/063} {\bibfield  {journal} {\bibinfo
  {journal} {JHEP}\ }\textbf {\bibinfo {volume} {04}},\ \bibinfo {pages} {063}
  (\bibinfo {year} {2008})},\ \Eprint {http://arxiv.org/abs/0802.1189}
  {arXiv:0802.1189 [hep-ph]} \BibitemShut {NoStop}%
\bibitem [{\citenamefont {Cacciari}\ \emph {et~al.}(2012)\citenamefont
  {Cacciari}, \citenamefont {Salam},\ and\ \citenamefont
  {Soyez}}]{Cacciari:2011ma}%
  \BibitemOpen
  \bibfield  {author} {\bibinfo {author} {\bibfnamefont {M.}~\bibnamefont
  {Cacciari}}, \bibinfo {author} {\bibfnamefont {G.~P.}\ \bibnamefont {Salam}},
  \ and\ \bibinfo {author} {\bibfnamefont {G.}~\bibnamefont {Soyez}},\ }\href
  {\doibase 10.1140/epjc/s10052-012-1896-2} {\bibfield  {journal} {\bibinfo
  {journal} {Eur. Phys. J.}\ }\textbf {\bibinfo {volume} {C72}},\ \bibinfo
  {pages} {1896} (\bibinfo {year} {2012})},\ \Eprint
  {http://arxiv.org/abs/1111.6097} {arXiv:1111.6097 [hep-ph]} \BibitemShut
  {NoStop}%
\bibitem [{\citenamefont {{ATLAS and CMS
  Collaborations}}(2019)}]{Atlas:2019qfx}%
  \BibitemOpen
  \bibfield  {author} {\bibinfo {author} {\bibnamefont {{ATLAS and CMS
  Collaborations}}} (\bibinfo {collaboration} {ATLAS, CMS}),\ }\href {\doibase
  10.23731/CYRM-2019-007.Addendum} {\bibfield  {journal} {\bibinfo  {journal}
  {CERN Yellow Rep. Monogr.}\ }\textbf {\bibinfo {volume} {7}},\ \bibinfo
  {pages} {Addendum} (\bibinfo {year} {2019})},\ \Eprint
  {http://arxiv.org/abs/1902.10229} {arXiv:1902.10229 [hep-ex]} \BibitemShut
  {NoStop}%
\bibitem [{\citenamefont {Berta}\ \emph {et~al.}(2014)\citenamefont {Berta},
  \citenamefont {Spousta}, \citenamefont {Miller},\ and\ \citenamefont
  {Leitner}}]{Berta:2014eza}%
  \BibitemOpen
  \bibfield  {author} {\bibinfo {author} {\bibfnamefont {P.}~\bibnamefont
  {Berta}}, \bibinfo {author} {\bibfnamefont {M.}~\bibnamefont {Spousta}},
  \bibinfo {author} {\bibfnamefont {D.~W.}\ \bibnamefont {Miller}}, \ and\
  \bibinfo {author} {\bibfnamefont {R.}~\bibnamefont {Leitner}},\ }\href
  {\doibase 10.1007/JHEP06(2014)092} {\bibfield  {journal} {\bibinfo  {journal}
  {JHEP}\ }\textbf {\bibinfo {volume} {06}},\ \bibinfo {pages} {092} (\bibinfo
  {year} {2014})},\ \Eprint {http://arxiv.org/abs/1403.3108} {arXiv:1403.3108
  [hep-ex]} \BibitemShut {NoStop}%
\bibitem [{\citenamefont {Cacciari}\ and\ \citenamefont
  {Salam}(2008)}]{Cacciari:2007fd}%
  \BibitemOpen
  \bibfield  {author} {\bibinfo {author} {\bibfnamefont {M.}~\bibnamefont
  {Cacciari}}\ and\ \bibinfo {author} {\bibfnamefont {G.~P.}\ \bibnamefont
  {Salam}},\ }\href {\doibase 10.1016/j.physletb.2007.09.077} {\bibfield
  {journal} {\bibinfo  {journal} {Phys.\ Lett.\ B}\ }\textbf {\bibinfo {volume}
  {659}},\ \bibinfo {pages} {119} (\bibinfo {year} {2008})},\ \Eprint
  {http://arxiv.org/abs/0707.1378} {arXiv:0707.1378 [hep-ph]} \BibitemShut
  {NoStop}%
\bibitem [{\citenamefont {Haake}\ and\ \citenamefont
  {Loizides}(2019)}]{Haake:2018hqn}%
  \BibitemOpen
  \bibfield  {author} {\bibinfo {author} {\bibfnamefont {R.}~\bibnamefont
  {Haake}}\ and\ \bibinfo {author} {\bibfnamefont {C.}~\bibnamefont
  {Loizides}},\ }\href {\doibase 10.1103/PhysRevC.99.064904} {\bibfield
  {journal} {\bibinfo  {journal} {Phys. Rev. C}\ }\textbf {\bibinfo {volume}
  {99}},\ \bibinfo {pages} {064904} (\bibinfo {year} {2019})},\ \Eprint
  {http://arxiv.org/abs/1810.06324} {arXiv:1810.06324 [nucl-ex]} \BibitemShut
  {NoStop}%
\bibitem [{\citenamefont {Mehtar-Tani}\ \emph {et~al.}(2019)\citenamefont
  {Mehtar-Tani}, \citenamefont {Soto-Ontoso},\ and\ \citenamefont
  {Verweij}}]{Yacine:2019ycj}%
  \BibitemOpen
  \bibfield  {author} {\bibinfo {author} {\bibfnamefont {Y.}~\bibnamefont
  {Mehtar-Tani}}, \bibinfo {author} {\bibfnamefont {A.}~\bibnamefont
  {Soto-Ontoso}}, \ and\ \bibinfo {author} {\bibfnamefont {M.}~\bibnamefont
  {Verweij}},\ }\href {\doibase 10.1103/PhysRevD.100.114023} {\bibfield
  {journal} {\bibinfo  {journal} {Phys. Rev. D}\ }\textbf {\bibinfo {volume}
  {100}},\ \bibinfo {pages} {114023} (\bibinfo {year} {2019})},\ \Eprint
  {http://arxiv.org/abs/1904.12815} {arXiv:1904.12815 [hep-ph]} \BibitemShut
  {NoStop}%
\bibitem [{\citenamefont {Berta}\ \emph {et~al.}(2019)\citenamefont {Berta},
  \citenamefont {Masetti}, \citenamefont {Miller},\ and\ \citenamefont
  {Spousta}}]{Berta:2019hnj}%
  \BibitemOpen
  \bibfield  {author} {\bibinfo {author} {\bibfnamefont {P.}~\bibnamefont
  {Berta}}, \bibinfo {author} {\bibfnamefont {L.}~\bibnamefont {Masetti}},
  \bibinfo {author} {\bibfnamefont {D.}~\bibnamefont {Miller}}, \ and\ \bibinfo
  {author} {\bibfnamefont {M.}~\bibnamefont {Spousta}},\ }\href {\doibase
  10.1007/JHEP08(2019)175} {\bibfield  {journal} {\bibinfo  {journal} {JHEP}\
  }\textbf {\bibinfo {volume} {08}},\ \bibinfo {pages} {175} (\bibinfo {year}
  {2019})},\ \Eprint {http://arxiv.org/abs/1905.03470} {arXiv:1905.03470
  [hep-ph]} \BibitemShut {NoStop}%
\bibitem [{\citenamefont {Andrews}\ \emph {et~al.}(2020)\citenamefont {Andrews}
  \emph {et~al.}}]{Andrews:2018jcm}%
  \BibitemOpen
  \bibfield  {author} {\bibinfo {author} {\bibfnamefont {H.~A.}\ \bibnamefont
  {Andrews}} \emph {et~al.},\ }\href {\doibase 10.1088/1361-6471/ab7cbc}
  {\bibfield  {journal} {\bibinfo  {journal} {J. Phys. G}\ }\textbf {\bibinfo
  {volume} {47}},\ \bibinfo {pages} {065102} (\bibinfo {year} {2020})},\
  \Eprint {http://arxiv.org/abs/1808.03689} {arXiv:1808.03689 [hep-ph]}
  \BibitemShut {NoStop}%
\bibitem [{fco()}]{fcontrib}%
  \BibitemOpen
  \href@noop {} {\enquote {\bibinfo {title} {Fastjet contrib},}\ }\bibinfo
  {howpublished} {\url{http://fastjet.hepforge.org/contrib/}}\BibitemShut
  {NoStop}%
\end{thebibliography}%

\end{document}